\documentclass[twocolumn,english,aps,prb,floatfix,superscriptaddress,showpacs,amsfonts,amssymb]{revtex4}
\usepackage[T1]{fontenc}
\usepackage[latin1]{inputenc}
\usepackage{amsmath}
\usepackage{amssymb}

\makeatletter

\usepackage{graphicx}
\usepackage{amscd}
\usepackage{bm}

\usepackage{babel}
\makeatother
\begin{document}

\title{Localization effects and inelastic scattering in disordered heavy electrons}

\author{M. C. O. Aguiar}

\affiliation{Instituto de F\'{\i}sica Gleb Wataghin, Unicamp, C.P. 6165, Campinas, SP 13083-970,
Brazil}

\author{E.~Miranda}

\affiliation{Instituto de F\'{\i}sica Gleb Wataghin, Unicamp, C.P. 6165, Campinas, SP 13083-970,
Brazil}

\author{V. Dobrosavljevi\'{c}}

\affiliation{Department of Physics and National High Magnetic Field Laboratory, Florida State
University, Tallahassee, FL 32306}

\date{\today{}}

\begin{abstract}
We study ground state and finite temperature properties of disordered heavy fermion
metals by using a generalization of dynamical mean field theory which incorporates
Anderson localization effects. The emergence of a non-Fermi liquid metallic behavior
even at moderate disorder is shown to be a universal phenomenon resulting from local
density of states fluctuations. This behavior is found to have a character of an
electronic Griffiths phase, and can be thought of as a precursor of Anderson localization
in a strongly correlated host. The temperature dependence of the conducting properties
of the system reveal a non-trivial interplay between disorder and inelastic processes,
which are reminiscent of the Mooij correlations observed in many disordered metals. 
\end{abstract}


\maketitle

\section{Introduction}

The interplay of disorder and strong correlations remains one of the least understood
topics of contemporary condensed matter physics. These effects are believed to bear
relevance to many problems that have attracted recent attention, such as the metal-insulator
transition (MIT) in two-dimensional electron systems.\cite{kravchenkormp} Disorder
effects are also likely to be important for the understanding of the puzzling non-Fermi
liquid (NFL) behavior of several heavy fermion compounds. \cite{stewartNFL} In some
of these systems, impurities seem to play only a subsidiary role: the explanation
for the anomalous behavior is more likely to be found in the physics of quantum criticality,\cite{lohneysenetal,grosche1,schroederetal,mathuretal,rosch}
even though a complete description is still lacking.\cite{hertz,japiassuetal,millis,sietal}
However, in other heavy fermion systems, disorder seems to play a more essential
role and seems to be at the origin of the NFL behavior.\cite{andrakastewart,bernaletal,taniguchietal,tabataetal}

Several attempts have been made to address theoretically the role of
disorder in heavy fermion compounds (see an overview below). Many
experimental results can be described within the so-called Kondo
disorder model (KDM)\cite{bernaletal} or, equivalently, the dynamical
mean field theory (DMFT) of disordered Kondo/Anderson
lattices,\cite{mirandavladgabi2,mirandavladgabi1} at least above the
lowest temperatures. Essential to this description is the
consideration of the full distribution of local Kondo temperatures
$T_{K}$. For sufficient disorder, it has a large weight as
$T_{K}\rightarrow 0$ describing the presence of dilute low-$T_{K}$
spins that dominate the thermodynamic and transport properties.
However, the KDM/DMFT predictions relied on a fine tuning of the bare
disorder that suggested that a more accurate microscopic foundation
was necessary. Conspicuously missing in this scheme were fluctuations
in the conduction electron local density of states (DOS), a quantity
that is crucial for the determination of $T_{K}$.
This was remedied by two of us through a generalization
of the DMFT that incorporates such Anderson localization effects while keeping its
local treatment of correlations.\cite{mirandavlad1,mirandavlad2} 
As a result, the question of the extreme sensitivity to the bare disorder 
was solved.
One important result of this study is the emergence of a power-law
distribution of Kondo temperatures $P\left(T_{K}\right)\propto
T_{K}^{\alpha -1}$, where $\alpha $ depends continuously on the
strength of disorder $W$ and decreases as $W$ is increased. As the
distribution becomes more singular, several thermodynamic quantities
become divergent, in a manner characteristic of Griffiths
phases.\cite{griffiths} The system eventually localizes at a critical
disorder strength $W_{MIT}$. Since there is no magnetic phase
transition this has been dubbed an electronic Griffiths phase.

This initial work\cite{mirandavlad1,mirandavlad2} employed the slave boson large-N
theory\cite{readnewns2,colemanlong} to solve the auxiliary single-impurity problems
posed by the method. While versatile, powerful and yet computationally cheap, this
approach presents some disadvantages, the main one being the difficulty of working
at finite temperatures. It should be reminded that, since we deal with wide distributions
of $T_{K}$, we need to be able to describe well the full crossover from $T\ll T_{K}$
to $T\gg T_{K}$, which is not possible with the slave boson large-N treatment. In
particular, conspicuously missing are inelastic scattering processes. Besides, though
giving a good description of the low-energy Fermi-liquid regime of the single-impurity
problem, this treatment does not incorporate high- and intermediate-energy incoherent
processes. Another impurity solver is therefore needed to assess the importance of
these intrinsically finite-$T$ and finite-energy features. A particularly useful
method, able to fill this gap at a reasonable computational cost, is second order
perturbation theory in $U$. We have used this method to both complement and cross-check
the slave boson results.

Besides the results of the initial
work,\cite{mirandavlad1,mirandavlad2} which have been confirmed by
both methods, some of our main conclusions are: (\emph{i}) there is a
subtle interplay between conduction and f-electron site disorder that
leads to a surprising \emph{non-monotonic} dependence of the
conducting properties on disorder, a feature that is likely unique to
Kondo/Anderson as opposed to Hubbard models; (\emph{ii}) localization
effects are essential for the determination of the distribution of
Kondo temperatures 
and a KDM/DMFT description is clearly insufficient,
especially if one starts from an experimentally measured discrete
distribution; (\emph{iii}) the interplay between disorder and
inelastic processes can lead to a temperature dependence of the
conducting properties that is reminiscent of the ones found by Mooij
and others in several strongly correlated disordered
metals.\cite{mooij}

This paper is organized as follows. We review the disorder-based mechanisms of non-Fermi
liquid behavior in the next subsection. Section~\ref{sec:The-model} describes the
model of disordered Anderson lattices we studied and the methods we employed to solve
it. Section~\ref{sec:Slave-boson} focuses on the detailed results obtained within
the slave boson large-N method. This expands considerably on the previously published
results.\cite{mirandavlad1,mirandavlad2} In Section~\ref{sec:Perturbation-theory},
we show the results obtained with perturbation theory. Finally, we wrap up with a
general discussion of the strengths and limitations of this study and point out possible
future directions in Section~\ref{sec:Discussion-and-conclusions}. Some details
of the computational procedures are given in an Appendix.

\subsection{Brief overview of disorder-based mechanisms of non-Fermi liquid behavior}

\subsubsection{Kondo-disorder models and the electronic Griffiths
phase}
\label{kdmandloc}

The KDM was proposed early on to account for the temperature dependence of the Cu
nuclear magnetic resonance (NMR) line widths in UCu$_{5-x}$Pd$_{x}$ ($x=0.5-1$).\cite{bernaletal}
It assumed that disorder in a heavy fermion system generates random spatial fluctuations
of the exchange coupling constant $J$ between local moments and conduction electrons
(the Kondo coupling). Each local moment was assumed to undergo the Kondo effect in
a manner that is completely uncorrelated with the others and each with a characteristic
energy scale, its Kondo temperature $T_{Kj}$. Even narrow Kondo coupling distributions
lead to a wide distribution of Kondo temperatures due to the latter's exponential
dependence on the former. As a result, at low temperatures, many spins are quenched
while a few percent remain unquenched and dominate, giving rise to singular, NFL
thermodynamic properties (specific heat and magnetic susceptibility). The NMR results
in UCu$_{5-x}$Pd$_{x}$ ($x=0.5-1$) are well described within this picture if the
distribution function $P\left(T_{K}\right)$ is such that $P\left(T_{K}\rightarrow 0\right)\rightarrow \mathrm{const.}$\cite{bernaletal}
The KDM gained a natural theoretical setting within the DMFT\cite{georgesrmp} of
a disordered Anderson/Kondo lattice.\cite{mirandavladgabi2,mirandavladgabi1} In
this approach, each conduction electron site exchanges single particle excitations
with an average {}``cavity'' bath, which is in turn self-consistently determined.
This treatment becomes exact in the limit of infinite dimensionality and is the natural
generalization of the Curie-Weiss mean field theory of magnets to a fermionic system.
Its treatment of disorder is equivalent to the well-known coherent potential approximation
(CPA).\cite{economou} In an Anderson lattice description, the localized f-electron
is hybridized with its adjacent conduction electron orbital and spatial fluctuations
are preserved through the random distribution of hybridization strengths. The local
moments are no longer independent since their distribution self-consistently determines
the cavity bath. Besides showing that the KDM corresponds to a rigorous limit of
a microscopic Hamiltonian, the DMFT enabled the calculation of other properties such
as the resistivity\cite{mirandavladgabi1,mirandavladgabi2}, dynamic magnetic susceptibility,\cite{mirandavladgabi1}
optical conductivity\cite{chatto} and magneto-resistance,\cite{chatto2} with good
agreement with experiments. The non-Fermi liquid behavior of these quantities hinged
on the condition that $P\left(T_{K}\rightarrow 0\right)\rightarrow \mathrm{const.}$
Subsequent experiments of muon spin rotation\cite{dougetal3} and NMR in high fields\cite{buttgenetal}
showed some inconsistencies with the KDM/DMFT, suggesting that inter-site correlations,
which are absent from that approach, may play a crucial role at the lowest temperatures.
However, annealing studies have further emphasized that the consideration of disorder
effects is indispensable.\cite{boothetal2}

Despite its success, the KDM/DMFT description suffered from a basic
deficiency, which can be ascribed to its extreme sensitivity to the
bare disorder distribution. Indeed, the connection between the
distribution of $J$ and the distribution of Kondo temperatures is too
rigid and a proper description always relies on fine tuning. In
particular, discrete distributions of bare parameters can never
generate a $P(T_K)$ such that $P\left(T_{K}\rightarrow
0\right)\rightarrow \mathrm{const.}$. Likewise, power law
distributions of $T_K$ are often necessary and it is not clear how
they can be obtained within the KDM/DMFT.

More recently, two of us have pointed out that this can be solved
through the inclusion of localization effects. From a basic
theoretical point of view, the importance of this modification is
unquestionable. Disorder scatters the conduction electrons giving rise
to spatial fluctuations in their wave function amplitude. These
Anderson localization precursor effects in turn give rise to
fluctuations of the conduction electron local DOS. The Kondo
temperatures are exponential functions of the local DOS and will show
a wide distribution for mild disorder strengths, \emph{even in the
absence of fluctuations in $J$}.\cite{vladtedgabi} Besides, direct
experimental determination of the distribution of $J$ from x-ray
absorption fine-structure (XAFS) experiments in UCu$_{5-x}$Pd$_{x}$
have shown that additional conduction electron disorder is necessary
for the interpretation of the results within the
KDM/DMFT.\cite{boothetal,baueretal} Finally, the addition of
localization effects has proved to be just the necessary ingredient
for the elimination of the extreme sensitivity to the bare disorder
and for a much more universal description.

The average cavity bath of the DMFT, however, completely neglects DOS
fluctuations and a more general treatment is necessary.  Progress
could be made by means of the statistical dynamical mean field theory
(statDMFT), which incorporates the full distribution of the conduction
electron local DOS, while keeping the treatment of local correlations
already present in the DMFT.\cite{vladgabisdmft1,vladgabisdmft2} The
treatment involves solving a fully self-consistent loop: the
f-electron fluid gives rise to an effective disorder potential for the
conduction electrons, while the latter's DOS fluctuations determine
the distribution of Kondo temperatures. We enumerate the main
conclusions of our analysis:\cite{mirandavlad1,mirandavlad2}

\begin{itemize}
\item \emph{Universality:} The distributions of several physical
quantities (Kondo temperatures, local DOS of f- and conduction
electrons, scattering T-matrices) assume a universal log-normal form
for weak to moderate disorder, irrespective of the form of the bare
distribution. We have verified this for Gaussian, square and discrete
bare distributions of conduction electron on-site energies and
hybridization strengths. This is in contrast to the KDM/DMFT results
and reflects the mixing of many different sites connected by the
extended conduction electron wave function.
\item \emph{Electronic Griffiths phase}: Increasing the disorder generates wide $T_{K}$
distributions, leading to a Griffiths phase with non-Fermi liquid behavior. This
Griffiths phase is not tied to any magnetic phase transition but is electronic in
origin: it is generated by the precursors to the Anderson localization transition. 
\item \emph{Metal-insulator transition}: There is an Anderson-type localization transition
at a critical value of the conduction electron diagonal disorder. 
\item \emph{Non-monotonic conductivity as a function of the disorder}: The typical conduction
electron DOS, which vanishes at the localization transition and serves as a measure
of the conducting properties shows a counter-intuitive non-monotonic behavior as
a function of disorder for a wide range of fillings. This surprising feature originates
in the interplay between the bare and the f-electron disorder potentials and is tied
to the proximity to the Kondo insulating (pseudo-)gap. 
\end{itemize}

\subsubsection{Magnetic Griffiths phase scenario}

An alternative theoretical scenario for disorder-induced non-Fermi liquid behavior
is the magnetic Griffiths phase.\cite{castronetoetal1,andradeetal,castronetojones}
In the vicinity of magnetic phase transitions disorder fluctuations induce rare regions
with an enhanced local critical temperature. These large clusters are ordered on
the scale of the correlation length and act as effective spins. Though rare in occurrence
they carry a considerable amount of magnetic entropy and the overall effect is the
appearance of singular, thermodynamic responses. This magnetic Griffiths phase picture
has been advocated as a source of NFL behavior in disordered heavy fermion systems.\cite{andradeetal}
However, very recent results seem to point to several difficulties encountered when
this scenario is applied to experimental systems, as follows.

\emph{The entropy problem.} The amount of magnetic entropy observed experimentally
in most disordered heavy fermion systems seems much too high to be compatible with
the magnetic Griffiths phase picture. Taking the measured specific heat of, say,
UCu$_{5-x}$Pd$_{x}$, we estimate that about 5\% of the sample would have to participate
in the spin-$\frac{1}{2}$ clusters. This implies an average cluster separation of
$2-3$ lattice constants, ruling out cluster sizes exceeding this distance. These
small clusters suggest instead that the {}``unquenched'' localized moments of the
KDM/DMFT or the electronic Griffiths phase offer a much more natural explanation.

\emph{Effects of dissipation.} Other important limitations of the magnetic Griffiths
phase scheme have also been emphasized in recent work by Millis, Morr, and Schmalian,\cite{millismorrschm}
who have carefully examined the effects of dissipation caused by the metallic bath.
This work suggests that the dissipation caused by itinerant electrons is so pronounced
that quantum tunneling of even moderately sized magnetic clusters will be suppressed.
Although the emergence of a magnetic Griffiths phase is a well established phenomenon
in disordered \emph{insulating} magnets, this result seems to bring into question
its relevance to itinerant systems.

\subsubsection{Spin glass precursors}

Finally, another possibility is to invoke the proximity to a spin glass quantum phase
transition. Several theoretical schemes predicting non-Fermi liquid behavior in the
vicinity of a spin-glass quantum critical point have been proposed.\cite{sachdevreadopper,senguptageorges,grempelmarcelo1}
Spin glass phases have been identified in the phase diagram of some heavy fermion
alloys (UCu$_{5-x}$Pd$_{x}$, for $x>1.5$)\cite{vollmeretal} and structurally
disordered compounds (URh$_{2}$Ge$_{2}$).\cite{sullowetal} More interestingly,
evidence of glassy dynamics \emph{in the absence of freezing} at very low temperatures
has been seen in UCu$_{5-x}$Pd$_{x}$ ($x=1$, $1.5$)\cite{dougetal} and Ce(Ru$_{0.5}$Rh$_{0.5}$)$_{2}$Si$_{2}$,\cite{yamamotoetal,tabataetal}
with conflicting results pointing to a very low freezing temperature in UCu$_{3.5}$Pd$_{1.5}$.\cite{vollmeretal}
These experiments seem to suggest that, if there is a true spin glass transition,
freezing temperatures are \emph{strongly suppressed} in a wide portion of the phase
diagram.

\section{The model and its solution}

\label{sec:The-model}

\subsection{The statistical dynamical mean field theory}

A simplified Hamiltonian capable of capturing the essential physics of disordered
metals with localized moments is provided by a disordered Anderson lattice\begin{equation}
H=H_{c}+H_{f}+H_{hyb},\label{eq:hamiltonian}\end{equation}
 where\begin{eqnarray}
H_{c} & = & -t\sum _{\left\langle ij\right\rangle \sigma }\left(c_{i\sigma }^{\dagger }c_{j\sigma }+\mathrm{H.}\, \mathrm{c.}\right)+\sum _{j\sigma }(\epsilon _{j}-\mu)c_{j\sigma }^{\dagger }c_{j\sigma };\label{eq:hamconduction}\\
H_{f} & = & \sum _{j\sigma }E_{f}f_{j\sigma }^{\dagger }f_{j\sigma }+U\sum _{j}f_{j\uparrow }^{\dagger }f_{j\uparrow }f_{j\downarrow }^{\dagger }f_{j\downarrow };\label{eq:hamlocal}\\
H_{hyb} & = & \sum _{j\sigma }\left(V_{j}f_{j\sigma }^{\dagger }c_{j\sigma }+\mathrm{H.}\, \mathrm{c.}\right).\label{eq:hamhybrid}
\end{eqnarray}
In Eqs.~(\ref{eq:hamconduction}-\ref{eq:hamhybrid}), $c_{j\sigma }$
($f_{j\sigma }$) annihilates a conduction (f-) electron on site $j$
with spin projection $\sigma $, $t$ is the nearest neighbor hopping
amplitude, $\mu$ is the chemical potential, $U$ is the f-site Coulomb
repulsion, $E_f$ is the f-energy level, and we introduce random
conduction electron on-site energies ($\epsilon _{j}$) and
hybridization matrix elements $V_{j}$.  These are chosen from given
distributions $P_{1}\left(\epsilon \right)$ and $P_{2}\left(V\right),$
taken to be either square or Gaussian, with width and standard
deviation $W$, respectively.  We have also studied discrete cases of
$P_{2}\left(V\right)$. There is a large degree of uncertainty as to a
realistic model of disorder for heavy fermion alloys. A rather
thorough study of the local f-site environment in the alloys
UCu$_{5-x}$Pd$_{x}$ ($x=1$ and $0.5$) was carried out in
Refs.~\onlinecite{boothetal,baueretal} through XAFS experiments. These
authors were able to determine the amount of Pd/Cu site interchange as
well as the U-Cu bond length distributions. In order to accommodate
both types of fluctuations one must, in principle, allow for a
distribution of both hybridization strengths \emph{and} on-site
conduction electron energies. On the other hand, when the local
moments are randomly replaced by non-magnetic elements (the so-called
{}``Kondo holes''), spatial fluctuations of $E_{f}$ should also be
included.\cite{mirandavladgabi1} Throughout the paper, we use the half
bandwidth $D$ as energy unit. For the Bethe lattice (to be introduced
later), $D=2\sqrt{2}t$ if the coordination number is 3. This should be
contrasted to Ref.~\onlinecite{mirandavlad1}, where $W$ is measured in
units of $t$. Note also that the f-level energy $E_f$ is always
measured \emph{relative to the chemical potential}. This is to ensure
that, by making it negative and large enough in absolute value, we
always work in the local moment/Kondo regime.

We worked within the framework of the statDMFT.\cite{vladgabisdmft1,vladgabisdmft2}
This treatment is able to incorporate both strong local correlations and Anderson
localization effects in a fully self-consistent fashion. Although the method has
been described before in the context of the disordered Hubbard model,\cite{vladgabisdmft2}
we will briefly review it with the dual goal of setting the notation and extending
it to the disordered Anderson lattice. It starts by focusing on a generic unit cell
$j$ of the lattice, containing an f-site and its adjoining conduction electron Wannier
state, and writing its effective action in imaginary time as \begin{eqnarray}
S_{eff}\left(j\right) & = & S_{c}\left(j\right)+S_{f}\left(j\right)+S_{hyb}\left(j\right),\label{eq:siteaction}\\
S_{c}\left(j\right) & = & \sum _{\sigma }\int _{0}^{\beta }d\tau \int _{0}^{\beta }d\tau ^{\prime }c_{j\sigma }^{\dagger }\left(\tau \right)\left[\delta \left(\tau -\tau ^{\prime }\right)\right.\nonumber \\
 & \times  & \left.\left(\partial _{\tau }+\epsilon _{j}-\mu \right)+\Delta _{cj}\left(\tau -\tau ^{\prime }\right)\right]c_{j\sigma }\left(\tau ^{\prime }\right),\label{eq:siteactionc}\\
S_{f}\left(j\right) & = & \int _{0}^{\beta }d\tau \left[\sum _{\sigma }\left(\partial _{\tau }+E_{f} \right)f_{j\sigma }^{\dagger }\left(\tau \right)f_{j\sigma }\left(\tau \right)\right.\nonumber \\
 & + & \left.Un_{fj\uparrow }\left(\tau \right)n_{fj\downarrow }\left(\tau \right)\hspace {-2.7cm}\phantom {\sum _{\sigma }\left(\partial _{\tau }+E_{f}-\mu \right)}\right],\label{eq:siteactionf}\\
S_{hyb}\left(j\right) & = & \sum _{\sigma }\int _{0}^{\beta }d\tau \left[V_{j}f_{j\sigma }^{\dagger }\left(\tau \right)c_{j\sigma }\left(\tau \right)+\mathrm{H.}\, \mathrm{c.}\right],\label{eq:siteactionhyb}
\end{eqnarray}
 where $n_{fj\sigma }=f_{j\sigma }^{\dagger }f_{j\sigma }$. In writing Eqs.~(\ref{eq:siteaction}-\ref{eq:siteactionhyb}),
a simplification has been made of retaining only quadratic contributions in fermionic
fields after integrating out the other sites (besides the instantaneous Hubbard term),
much like the usual dynamical mean field theory.\cite{georgesrmp} The bath (or {}``cavity'')
function $\Delta _{cj}\left(\tau \right)$ in Eq.~(\ref{eq:siteactionc}) is given
by\begin{equation}
\Delta _{cj}\left(\tau \right)=t^{2}\sum _{l,m=1}^{z}G_{clm}^{\left(j\right)}\left(\tau \right),\label{eq:bath}\end{equation}
 where the sum extends over the $z$ nearest neighbors and \[
G_{clm}^{\left(j\right)}\left(\tau \right)=-\left\langle T\left[c_{m\sigma }\left(\tau \right)c_{l\sigma }^{\dagger }\left(0\right)\right]\right\rangle ^{\left(j\right)}\]
 is the Green's function for propagation from nearest neighbor site $l$ to nearest
neighbor site $m$, calculated with the site $j$ removed. Integrating out the remaining
conduction electron $c_{j\sigma }$, we get the effective action of an auxiliary
single-impurity Anderson model at each site $j$\begin{eqnarray}
S_{imp}\left(j\right) & = & \sum _{\sigma }\int _{0}^{\beta }d\tau \int _{0}^{\beta }d\tau ^{\prime }f_{j\sigma }^{\dagger }\left(\tau \right)\left[\delta \left(\tau -\tau ^{\prime }\right)\right.\nonumber \\
 &  & \left.\left(\delta _{\tau }+E_{f} \right)+\Delta _{fj}\left(\tau -\tau ^{\prime }\right)\right]f_{j\sigma }\left(\tau ^{\prime }\right)\nonumber \\
 &  & +\int _{0}^{\beta }d\tau Un_{fj\uparrow }\left(\tau \right)n_{fj\downarrow }\left(\tau \right),\label{eq:faction}
\end{eqnarray}
 where the hybridization function for the f-site is, in Matsubara frequency space,

\begin{equation}
\Delta _{fj}\left(i\omega _{n}\right)=\frac{V_{j}^{2}}{i\omega _{n}-\epsilon _{j}+\mu -\Delta _{cj}\left(i\omega _{n}\right)}.\label{eq:hybfunc}\end{equation}
 The solution of this impurity problem is the major difficulty in this treatment.
We implemented two different methods of solution, which will be expanded upon in
the next subsections. The aim is to calculate the local f-electron Green's function\begin{equation}
G_{fj}^{loc}\left(\tau \right)=-\left\langle T\left[f_{j\sigma }\left(\tau \right)f_{j\sigma }^{\dagger }\left(0\right)\right]\right\rangle ,\label{eq:gflocal}\end{equation}
 under the dynamics dictated by (\ref{eq:faction}). It is conveniently parameterized
by its self-energy\begin{equation}
G_{fj}^{loc}\left(i\omega _{n}\right)=\frac{1}{i\omega _{n}-E_{f} -\Delta _{fj}\left(i\omega _{n}\right)-\Sigma _{fj}\left(i\omega _{n}\right)}.\label{eq:gflocalmat}\end{equation}
 It is also convenient to define a local conduction electron Green's function\begin{equation}
G_{cj}^{loc}\left(\tau \right)=-\left\langle T\left[c_{j\sigma }\left(\tau \right)c_{j\sigma }^{\dagger }\left(0\right)\right]\right\rangle ,\label{eq:gclocal}\end{equation}
 such that\begin{equation}
G_{cj}^{loc}\left(i\omega _{n}\right)=\frac{1}{i\omega _{n}-\epsilon _{j}+\mu -\Delta _{cj}\left(i\omega _{n}\right)-\Phi _{j}\left(i\omega _{n}\right)},\label{eq:gclocalmat}\end{equation}
 where\begin{equation}
\Phi _{j}\left(i\omega _{n}\right)=\frac{V_{j}^{2}}{i\omega _{n}-E_{f} -\Sigma _{fj}\left(i\omega _{n}\right)}.\label{eq:phi}\end{equation}
 We note that the $\Phi _{j}\left(i\omega _{n}\right)$ function describes the local
scattering of the conduction electrons off the f-shell at site $j$, incorporating
information about both elastic and inelastic processes.

All the information a generic site $j$ has about the rest of the lattice is encoded
in the bath function (\ref{eq:bath}), which should be viewed as a functional of
the conduction electron lattice Green's function. Since a fully analytical treatment
is impossible, we have to solve the equations numerically. For this purpose, we formulated
the problem on a Bethe lattice, where things are considerably simplified as explained
in Ref.~\onlinecite{vladgabisdmft2}. In this case, nearest neighbors $l$ and $m$
become disconnected once $j$ is removed and only local Green's functions survive
\[
G_{clm}^{\left(j\right)}\left(i\omega _{n}\right)=\delta _{l,m}G_{cll}^{\left(j\right)}\left(i\omega _{n}\right)=\delta _{l,m}G_{cl}^{loc\left(j\right)}\left(i\omega _{n}\right).\]
 Finally, these last objects can be computed from an action at site $l$ in almost
all aspects identical to (\ref{eq:siteaction}-\ref{eq:siteactionhyb}), the only
difference now being that the bath function sum runs over the $z-1$ nearest neighbors
(here labeled by $m$) only\begin{equation}
\Delta _{cl}^{\left(j\right)}\left(i\omega _{n}\right)=t^{2}\sum _{m=1}^{z-1}G_{cm}^{loc\left(l\right)}\left(i\omega _{n}\right).\label{eq:bath2}\end{equation}
 Note that, on the right-hand side, we do not need to specify that site $j$ has
been excluded as the removal of $l$ completely disconnects sites labeled by $m$
from site $j$ (this property is specific to the Bethe lattice). The reappearance
of $G_{cm}^{loc\left(l\right)}\left(i\omega _{n}\right)$, whose distribution is
identical to that of $G_{cl}^{loc\left(j\right)}\left(i\omega _{n}\right)$ since
all sites are equivalent, closes the loop and establishes a recursive set of stochastic
equations. When the interaction is turned off ($U=0$), this treatment reduces to
the well-known self-consistent theory of localization,\cite{abouetal} here generalized
to a two-band lattice. When we take the coordination to infinity $z\rightarrow \infty $
keeping $\tilde{t}=t/\sqrt{z}=\mathrm{const.}$, our treatment reduces to the DMFT
of correlations and disorder.\cite{georgesrmp,mirandavladgabi1,mirandavladgabi2}
In the latter case, the disorder treatment is equivalent to the CPA,\cite{economou}
which has no localization transition.

A full solution of Eqs.~(\ref{eq:siteaction}-\ref{eq:bath2}) for given distributions
$P_{1}\left(\epsilon \right)$ and/or $P_{2}\left(V\right)$ involves solving an
ensemble of impurity problems self-consistently. Physically, the conduction electrons
propagate through a disordered lattice and scatter off conduction site potential
fluctuations as well as f-site resonances. These resonances, in turn, describe the
formation of localized moments, whose local Kondo temperatures fluctuate as well,
reflecting a disordered conduction sea environment. Complete statistical information
can in principle be obtained from the distributions of the various renormalized local
quantities. We stress that a random distribution of any bare parameter causes all
renormalized quantities to fluctuate as a result of the self-consistent nature of
our treatment. Therefore, even if we include only fluctuations in the conduction
sea through $P_{1}\left(\epsilon \right)$, \emph{a distribution of Kondo temperatures
ensue}s.\cite{vladtedgabi,zarandudvardi} This is easily seen from the approximate
formula for the Kondo temperature in the Kondo limit ($\sum _{\sigma }n_{fj\sigma }\approx 1$)\cite{hewson}\begin{equation}
T_{Kj}\approx D\exp \left(-\frac{1}{\rho _{j}J_{j}}\right),\label{eq:kondotemp}\end{equation}
 where $\rho _{j}$ is the local density of states (DOS) seen by the f-site\begin{equation}
\rho _{j}\approx \frac{\mathrm{Im}\left[\Delta _{fj}\left(0-i\eta \right)\right]}{\pi V_{j}^{2}}\label{eq:localrhoc}\end{equation}
 and $J_{j}$ is the Kondo coupling constant, given in the Kondo limit by\begin{equation}
J_{j}\approx 2V_{j}^{2}\left(\frac{1}{\left|E_{f}\right|}+\frac{1}{\left|E_{f}+U\right|}\right).\label{eq:kondoj}\end{equation}
 Even if $V_{j}$ is not random, the local DOS is, because of the denominator in
Eq.~(\ref{eq:hybfunc}). As a result of the strong exponential dependence in (\ref{eq:kondotemp}),
even mild localization effects can be strongly enhanced and should be seriously considered,
specially in disordered heavy fermion systems.

\subsection{The impurity solvers}

An important part of the method we employed is the solution of the impurity problems
posed by the ensemble of effective actions given by Eq.~(\ref{eq:faction}) and
its counterpart for a site with one nearest neighbor removed. We concentrated mostly
on two methods of solution, which we now briefly describe: the slave boson, large-N
based, mean field theory and second order perturbation theory in $U$. In order to
unclutter the notation, we drop in the next subsections the site index $j$ and the
superscript $loc$. Details of the numerical treatment are given in the Appendix.

\subsubsection{Slave boson mean field theory}

This method gives a good description of the low temperature, Fermi liquid regime
of the Anderson impurity problem in the limit $U\rightarrow \infty $ and is extensively
covered in the literature.\cite{readnewns2,colemanlong} Its main advantage is the
ability to capture the zero temperature fixed point correctly as well as the exponential
nature of the low energy scale. Its treatment of the self-energy, however, does not
incorporate inelastic processes to leading order. Besides, it has a spurious phase
transition at a finite temperature, where in reality there should be only a smooth
crossover. For these reasons, we confine it to the zero temperature limit, where
it is a useful guide. As applied to our problem, the method has been described in
Appendix D of Ref. \onlinecite{mirandavladgabi1} and we will merely state the results,
generalized to the Matsubara frequency axis and at $T=0$. The local f-electron Green's
function is given by\begin{eqnarray}
G_{f}\left(i\omega \right) & = & \frac{q}{i\omega -\epsilon _{f} -q\Delta _{f}\left(i\omega \right)}\label{eq:gflocsb}\\
 & \equiv  & qG_{f}^{qp}\left(i\omega \right),\label{eq:gfqp}
\end{eqnarray}
 where the last equality defines the local f-electron \emph{quasi-particle} Green's
function and the variational parameters $\epsilon _{f}$ (renormalized f-energy)
and $q$ (quasi-particle residue) are determined from the solution of the set of
equations

\begin{eqnarray}
\epsilon _{f}-E_{f}+\int _{-\infty }^{\infty }\frac{d\omega }{\pi }\Delta _{f}\left(i\omega \right)G_{f}^{qp}\left(i\omega \right) & = & 0,\label{eq:sb1}\\
q+\int _{-\infty }^{\infty }\frac{d\omega }{\pi }e^{i\omega \eta }G_{f}^{qp}\left(i\omega \right) & = & 1.\label{eq:sb2}
\end{eqnarray}
 Using\[
\int _{-\infty }^{\infty }\frac{d\omega }{\pi }e^{i\omega \eta }\mathrm{Im}\left[G_{f}^{qp}\left(i\omega \right)\right]=1,\]
 Eqs.~(\ref{eq:sb1}-\ref{eq:sb2}) simplify to\begin{eqnarray}
2\int _{0}^{\infty }\frac{d\omega }{\pi }\mathrm{Re}\left[\Delta _{f}\left(i\omega \right)G_{f}^{qp}\left(i\omega \right)\right] & = & E_{f}-\epsilon _{f},\label{eq:sb1b}\\
q+2\int _{0}^{\infty }\frac{d\omega }{\pi }\mathrm{Re}\left[G_{f}^{qp}\left(i\omega \right)\right] & = & 0.\label{eq:sb2b}
\end{eqnarray}
 Eq.~(\ref{eq:phi}) becomes in this approximation\begin{equation}
\Phi \left(i\omega \right)=\frac{qV^{2}}{i\omega -\epsilon _{f}}.\label{eq:phisb}\end{equation}

\subsubsection{Second order perturbation theory}

The perturbative solution of the single-impurity Anderson model with particle-hole
symmetry was thoroughly analyzed by Yamada and Yosida.\cite{yosidayamada1,yamada2,yosidayamada3,yamada4}
The series expansion in $U$ for physical quantities such as the specific heat and
the spin susceptibility converges very fast and even second order results can be
useful.\cite{zlatichorvatic} Extension of the perturbative treatment to the case
without particle-hole symmetry poses considerable difficulties. A particularly useful
proposal is the use of an interpolative self-energy which recovers the atomic ($V\rightarrow 0$)
and high frequency limits.\cite{ferreretal} Further improvements of the method were
later suggested.\cite{kajuetergabi,potthoffetal}

The procedure consists in defining an unperturbed f-electron Green's function\begin{equation}
G_{f}^{(0)}(i\omega _{n})=\frac{1}{i\omega _{n}+\widetilde{\mu }-\Delta _{f}(i\omega _{n})},\label{eq:fgreen0}\end{equation}
 with a new parameter $\widetilde{\mu }$ to be determined later, which vanishes
at particle-hole symmetry. The interacting Green's function is given in (\ref{eq:gflocalmat}).
The interpolative self-energy is\cite{kajuetergabi,potthoffetal}\begin{equation}
\Sigma _{f}(i\omega _{n})=Un+\frac{A\Sigma ^{(2)}(i\omega _{n})}{1-B\Sigma ^{(2)}(i\omega _{n})},\label{eq:fselfen}\end{equation}
 where\begin{equation}
n=T\sum _{\omega _{n}}e^{i\omega _{n}\eta }G_{f}(i\omega _{n}),\label{eq:nf}\end{equation}
 and\begin{equation}
\Sigma ^{(2)}(i\omega _{n})=-U^{2}\int _{0}^{\beta }d\tau e^{i\omega _{n}\tau }\left[G_{f}^{(0)}(\tau )\right]^{2}G_{f}^{(0)}(-\tau ).\label{eq:2ordselfen}\end{equation}
 The last equation (\ref{eq:2ordselfen}) is the usual second order diagram using
the unperturbed Green's function (\ref{eq:fgreen0}) for the internal lines. The
parameters $A$ and $B$ are determined by imposing the high frequency and atomic
limits, respectively, and are given by\cite{kajuetergabi}\begin{eqnarray}
A & = & \frac{n(1-n)}{n_{0}(1-n_{0})},\label{eq:aparam}\\
B & = & \frac{(1-n)U+E_{f}+\tilde{\mu }}{n_{0}(1-n_{0})U^{2}},\label{eq:bparam}
\end{eqnarray}

\noindent where

\begin{equation}
n_{0}=T\sum _{\omega _{n}}e^{i\omega _{n}\eta }G_{f}^{(0)}(i\omega _{n}).\label{eq:n0}\end{equation}
 Different schemes have been proposed in order to fix the free parameter $\widetilde{\mu }$.\cite{ferreretal,kajuetergabi,potthoffetal}
At zero temperature, one can ensure that the low energy Fermi liquid behavior is
obtained by imposing the Friedel sum rule.\cite{kajuetergabi} This procedure cannot
be easily generalized to finite temperatures, however. One option is to fix $\widetilde{\mu }$
at its zero temperature value even at finite temperatures. Alternatively, one can
require at any temperature\cite{ferreretal}\begin{equation}
n=n_{0},\label{eq:neqn0}\end{equation}
 which makes $A=1$. Finally, a third possibility is imposing $\widetilde{\mu }=\mu $.\cite{potthoffetal}
These three alternatives have been rather carefully compared in Ref.~\onlinecite{potthoffetal}
at $T=0$ and checked against exact diagonalization. The first two methods were shown
to be almost equivalent whereas the third one is inferior. Moreover, comparisons
at finite temperatures with Quantum Monte Carlo results confirmed the adequacy of
imposing Eq.~(\ref{eq:neqn0}).\cite{wegneretal} Specific applications to a clean
Anderson lattice model further corroborated this conclusion.\cite{meyernolting1,meyernolting2}
Thus, our results were based on imposing condition (\ref{eq:neqn0}). It should be
remembered, however, that the perturbative solution predicts a characteristic energy
scale that is \emph{quantitatively} incorrect at large values of $U$, since it is
unable to capture the correct exponential dependence. Nevertheless, for moderate
interactions, it still gives reasonable results. Within its limitations, this perturbative
scheme is a relatively flexible low-cost tool to tackle the impurity problem with
the great advantage of being able to naturally account for \emph{inelastic} processes.

We note that a direct comparison between the slave boson mean field theory results
and second order perturbation theory is not possible because the former is limited
to the $U\rightarrow \infty $ limit, which is obviously outside the region of validity
of the latter. The main interest of an analysis of both methods, however, resides
in the exploration of \emph{the importance of inelastic processes}, which are absent
in the slave boson mean field treatment.

\section{Slave boson mean field theory results}

\label{sec:Slave-boson}

We now present the results obtained at $T=0$ using the slave boson mean field theory
as an impurity solver. Most of our results were obtained for a uniform distribution
of on-site conduction electron energies\[
P_{1}\left(\epsilon \right)=\frac{1}{W};\, |\epsilon |\leq \frac{W}{2}.\]
 In Section~\ref{sub:f-electrons} we also show results for a discrete distribution
of hybridization strengths $P_{2}\left(V\right)$.

\subsection{Conduction electron typical density of states}

To understand the overall behavior as a function of disorder, it is instructive to
consider the transport properties of the conduction electrons. Since there are no
interactions among them in our model, their behavior is that of a disordered non-interacting
electron system. There are two sources of disorder, as can be seen in Eq.~(\ref{eq:gclocalmat}):
fluctuations of the local on-site energies $\epsilon _{j}$ and of the f-shell resonances
described by $\Phi _{j}\left(i\omega _{n}\right)$. They are not independent, however,
since they are inextricably tied by self-consistency. Their combined effect acts
to decrease the conduction electron mobility.

\subsubsection{Typical density of states:\protect \\
 an order parameter for localization}

A useful measure of this mobility is given by the \emph{typical} value of the local
escape rate. This is encoded in the imaginary part of the local conduction electron
Green's function (the local DOS) at zero frequency, $\rho _{cj}=\frac{1}{\pi }\mathrm{Im}\left[G_{cj}^{loc}\left(0-i\delta \right)\right]$.
We will, from now on, drop the superscript denoting the removal of a nearest neighbor
so as to lighten the notation. As shown originally by Anderson, the typical value
of the local DOS vanishes when the electrons are localized and can be viewed as an
order parameter for the localization transition.\cite{andersonloc} A convenient
way of accessing the typical value is furnished by the geometric average\begin{equation}
\rho _{c}^{typ}=\exp \left\{ \overline{\ln \rho _{cj}}\right\} ,\label{eq:rhoctyp}\end{equation}
 where the overbar denotes a disorder average. By contrast, the arithmetic average\begin{equation}
\rho _{c}^{av}=\overline{\rho _{cj}}\label{eq:rhocav}\end{equation}
 is finite at the transition. A thorough analysis of the critical behavior of the
local DOS distribution in the non-interacting Bethe lattice localization problem
was carried out in Ref.~\onlinecite{mirlinfyodorov}.

\begin{figure}
\begin{center}\includegraphics[  width=2.5in,
  keepaspectratio]{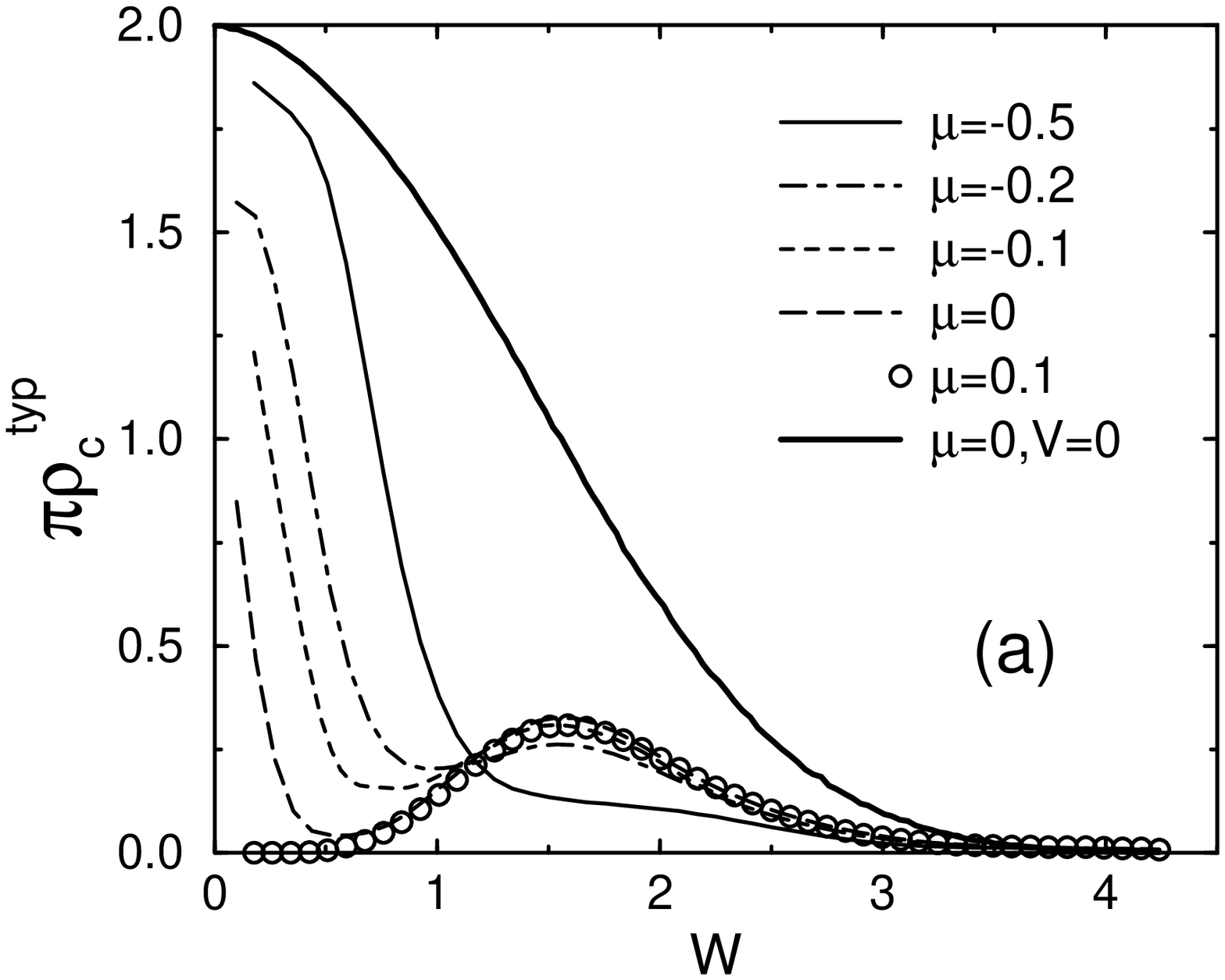}\end{center}

\begin{center}\includegraphics[  width=2.5in,
  keepaspectratio]{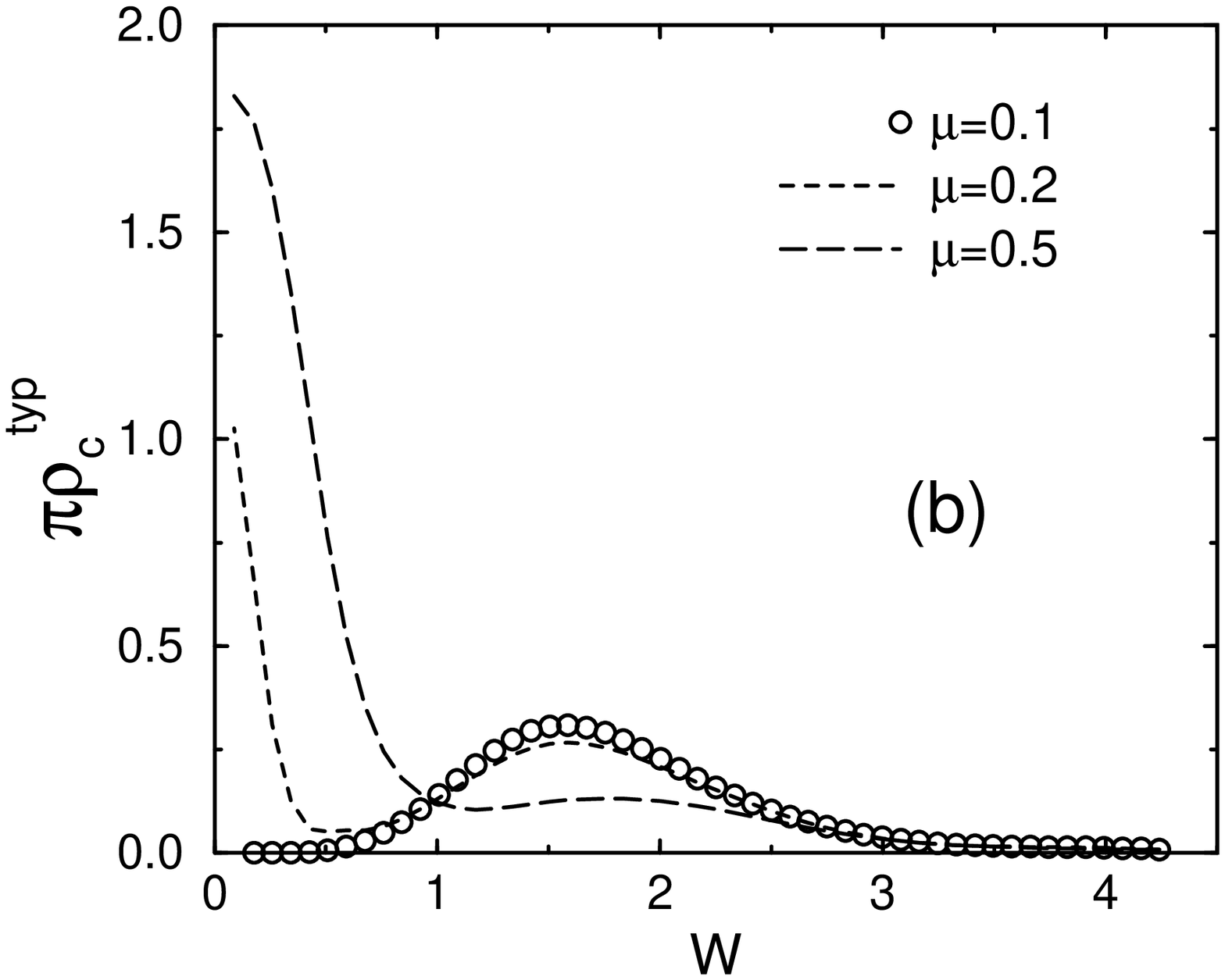}\end{center}

\caption{\label{cap:fig1}Typical conduction electron density of states as a function of
disorder strength for several values of the chemical potential $\mu $, using slave
boson mean field theory as the impurity solver. We used $E_{f}=-1$, $V=0.5$, except
for the thick solid line in (a), which is for $V=0.$}
\end{figure}

In Fig.~\ref{cap:fig1} we show the typical conduction electron DOS as a function
of disorder for several values of the chemical potential.

\subsubsection{Proximity to the Kondo insulator}

We can identify three qualitatively distinct behaviors.\cite{mirandavlad2} For $\mu \alt -0.4$
and for $\mu \agt 0.5$, $\rho _{c}^{typ}$ is a monotonically decreasing function
of disorder. For $\mu \approx 0.1,$ the clean system is a Kondo insulator\cite{aepplifisk,riseboroughKI}
and $\rho _{c}^{av}=\rho _{c}^{typ}=0$. As disorder is introduced in the Kondo insulator,
$\rho _{c}^{typ}$ initially increases, reaching a maximum at about $W\approx 1.5$,
after which it decreases monotonically. Finally, for $-0.3\alt \mu \alt 0.1$ and
$0.1\alt \mu \alt 0.4$ $\rho _{c}^{typ}$ initially decreases, passes through a
minimum around $W\approx 0.5-1$, then increases up to a maximum at about $W\approx 1.5$
and eventually becomes monotonically decreasing. For all values of $\mu $, the typical
DOS vanishes at a disorder-induced metal insulator transition (MIT) at $W_{MIT}\approx 4.5$.

These distinct behaviors can be traced back to how close the clean system is to the
Kondo insulator point\cite{aepplifisk,riseboroughKI} at $\mu \approx 0.1$.\cite{mirandavlad2}
If we start from the clean insulator, the introduction of disorder acts to create
states inside the gap, thus increasing the DOS at the chemical potential. This increase
continues until the gap is essentially washed out and the system becomes a bad metal.
After that, localization intervenes and $\rho _{c}^{typ}$ starts to decrease towards
the MIT.

For fillings close to but not at the Kondo insulator point, the clean system is a
heavy fermion metal at $T=0$. The f-resonances ($\Phi \left(\omega \right)$) coherently
scatter the conduction electrons creating a strongly renormalized Fermi liquid. In
the slave boson treatment, $\Phi \left(\omega \right)$ diverges at $\omega =\epsilon _{f}$,
see Eq.~(\ref{eq:phisb}), corresponding to the limit of unitary scattering, with
a maximally allowed phase shift $\delta =\pi /2$. We can view its value at the Fermi
level $\Phi \left(0\right)=-qV^{2}/\epsilon _{f}$ (which is real at $T=0)$ as an
effective potential coming from the f-electrons. The closer the system is to the
Kondo insulator, the larger the value of $\Phi \left(0\right)$, the insulator being
signaled by the divergence of this quantity (or equivalently by $\epsilon _{f}=0$).
The effect of disorder is to immediately start generating spatial fluctuations of
the f-resonances, with different phase shift values at the chemical potential. Proximity
to the insulator implies large, random, almost unitary scattering potentials. As
a result, metallic coherence is efficiently destroyed and the typical conduction
electron DOS is strongly suppressed.\cite{mirandavlad1,mirandavlad2} The important
role played by the unitary scatterers was emphasized in Refs.~\onlinecite{mirandavlad2}
and \onlinecite{mirandavlad1}, where the distribution $P\left(1/\left[\Phi \left(0\right)\right]\right)$
was directly computed and its weight at $1/\Phi \left(0\right)=0$ was shown to correlate
with the destruction of coherence.

There is another equivalent way of understanding these effects. For small dopings
away from the Kondo insulator, carriers are introduced at the \emph{edges} of the
valence or conduction bands defined by the Kondo insulator gap, which have a small
DOS (in the Bethe lattice, band edges have a square root shape as in three dimensions).
As has been known for a long time, a region of small DOS is particularly sensitive
to localization effects introduced by disorder.

As in the previous case, further increase of disorder acts to wash out the nearby
Kondo pseudo-gap and the behavior then becomes very similar to the disordered Kondo
insulator. We thus have a region with a rather non-intuitive increasing $\rho _{c}^{typ}$,
which can be ascribed to the proximity to the Kondo insulator fixed point. The behavior
at fillings well away from the Kondo insulator is much less influenced by the pseudo-gap,
see Fig.~\ref{cap:fig1} for $\mu =\pm 0.5$. Although there is a rapid initial
decrease of $\rho _{c}^{typ}$, followed by a much slower dependence, the typical
DOS does not exhibit the unconventional increase with disorder observed at other
fillings.

\subsubsection{Role of the hybridization strength}

\begin{figure}
\begin{center}\includegraphics[  width=2.5in,
  keepaspectratio]{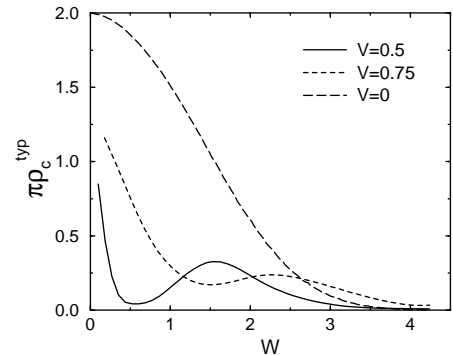}\end{center}

\caption{\label{cap:fig2}Typical conduction electron density of states as a function of
disorder strength for different values of hybridization $V$, using slave boson mean
field theory as the impurity solver ($E_{f}=-1$ and $\mu =0$).}
\end{figure}

It is interesting to note that the critical value of disorder for the MIT $W_{MIT}$
depends on the hybridization strength. In Fig.~\ref{cap:fig2}, we show the disorder
dependence of the typical conduction electron DOS for different values of $V$. There
is hardly any change in $W_{MIT}$ as we go from $V=0$ to $V=0.5$ (cf. also Fig.~\ref{cap:fig1}).
However, for $V=0.75$, the critical disorder strength is clearly enhanced. This
figure also illustrates the non-trivial nature of the self-consistency. Indeed, the
two types of disorder coming from fluctuations in $\epsilon _{j}$ and $\Phi _{j}$
are clearly not independent, since the \emph{addition} of f-site disorder as we turn
on $V$ from $0$ to $0.75$ acts to \emph{increase} the mobility for $W\agt 2.8$.
The self-consistently determined solutions of the impurity problems effectively help
\emph{screen} the conduction electron disorder. Note also how an increased value
of $V$ pushes the {}``dip-hump'' structure to higher values of disorder. Since
the Kondo insulating gap increases with the hybridization strength, this is consistent
with our explanation for the nature of this non-monotonic behavior.

\subsection{Distribution of the conduction electron local density of states}

\begin{figure}
\begin{center}\includegraphics[  width=2.25in,
  keepaspectratio]{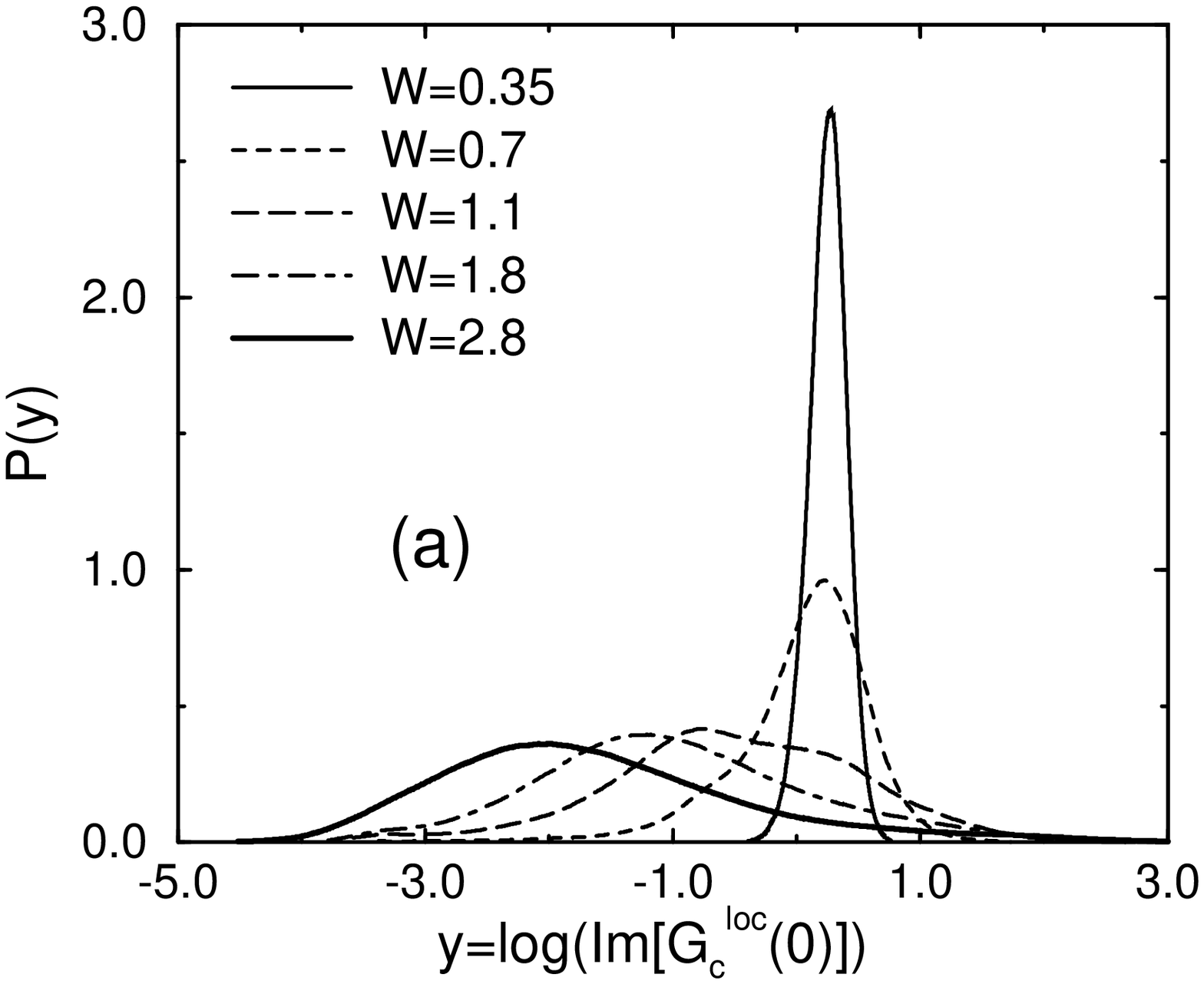}\end{center}

\begin{center}\includegraphics[  width=2.25in,
  keepaspectratio]{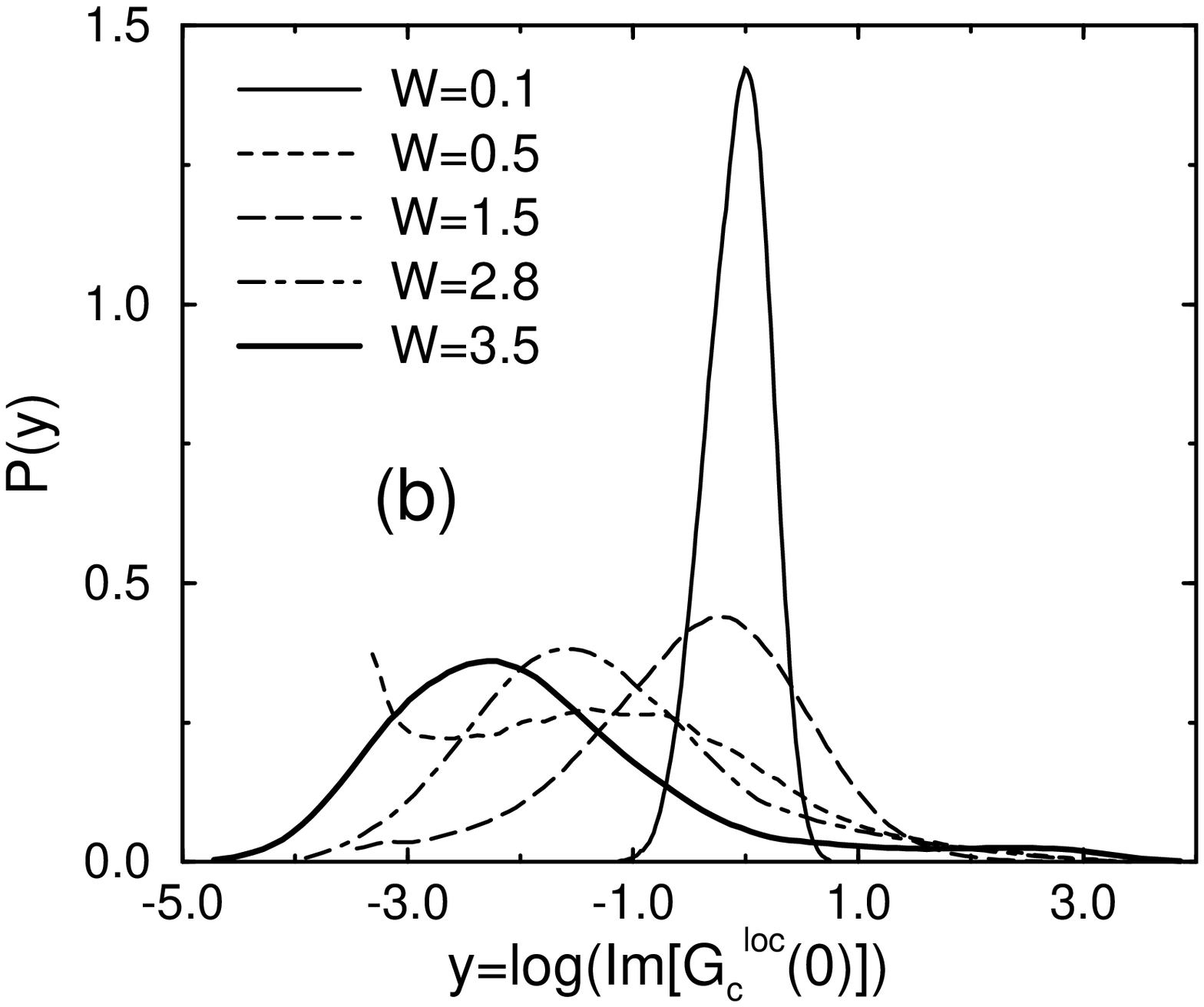}\end{center}

\begin{center}\includegraphics[  width=2.25in,
  keepaspectratio]{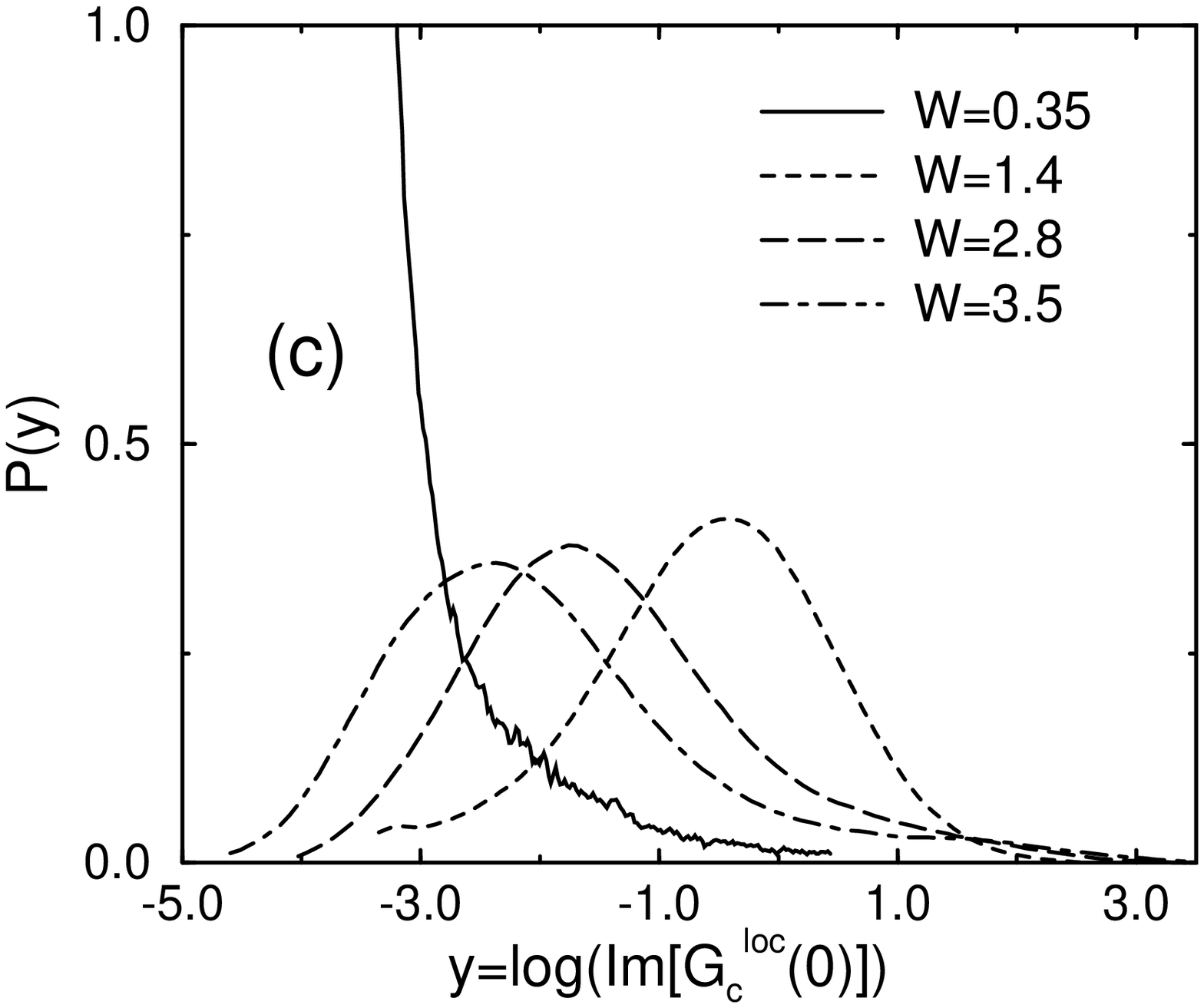}\end{center}

\caption{\label{cap:fig3}Distributions of the logarithm of $\mathrm{Im}\left[G_{c}^{loc}\left(0-i\delta \right)\right]$
for several values of disorder strength, using slave boson mean field theory as the
impurity solver: (a) $\mu =-0.5$, (b) $\mu =0$ and (c) $\mu =0.1$ ($V=0.5$, $E_{f}=-1$).}
\end{figure}

One of the great advantages of the present approach is the possibility of monitoring
complete distribution functions. Many of the features exhibited in Figs.~\ref{cap:fig1}
and \ref{cap:fig2} can be directly read off the distribution of $\pi \rho _{c}=\mathrm{Im}\left[G_{c}^{loc}\left(0-i\delta \right)\right]$.
We show this by plotting $P\left(\log \, \mathrm{Im}\left[G_{c}^{loc}\left(0-i\delta \right)\right]\right)$
for several disorder strengths and different chemical potential values in Fig.~\ref{cap:fig3}
(we use {}``log'' for the base 10 logarithm). It follows from the definition of
$\pi \rho _{c}^{typ}$, Eq.~(\ref{eq:rhoctyp}), that it is obtained by raising
10 to the power of the average of this distribution (we use powers of 10 for ease
of computation). For weak disorder in the metallic cases (Figs.~\ref{cap:fig3}a
and b), $P\left(\log \, \mathrm{Im}\left[G_{c}^{loc}\left(0-i\delta \right)\right]\right)$
is approximately Gaussian, \emph{even though the bare disorder is uniform}, a feature
shared by several physical quantities.\cite{mirandavlad1,mirandavlad2} This is due
to the presence of correlations between many distant lattice sites mediated by the
extended conduction electron wave function, which introduces a sort of averaging
effect. In the Kondo insulator case (Fig.~\ref{cap:fig3}c), however, the distribution
is \emph{not} Gaussian at weak disorder. Keeping in mind that $\rho _{c}=0$ ($\log \, \mathrm{Im}\left[G_{c}^{loc}\left(0-i\delta \right)\right]\rightarrow -\infty $)
in the clean Kondo insulator, it is clear that the introduction of weak disorder
has to generate weight at very small $\pi \rho _{c}$. Indeed, $P\left(\log \, \mathrm{Im}\left[G_{c}^{loc}\left(0-i\delta \right)\right]\right)$
shows a divergence at $\mathrm{Im}\left[G_{c}^{loc}\left(0-i\delta \right)\right]\approx 10^{-3}$
for $W=0.35$.

For large values of $W$ the distribution becomes extremely broad, spanning many
orders of magnitude. In the case $\mu =-0.5$ (Fig.~\ref{cap:fig3}a), corresponding
to a system well away from the Kondo insulator filling, the distribution broadens
and its maximum steadily shifts towards lower values as disorder is increased. This
is to be expected from the monotonic behavior of $\rho _{c}^{typ}$. Likewise, at
$\mu =0$, the non-monotonic behavior of the typical value is also clearly reflected
in $P\left(\log \, \mathrm{Im}\left[G_{c}^{loc}\left(0-i\delta \right)\right]\right)$
(see Fig.~\ref{cap:fig3}b and compare it to Fig.~\ref{cap:fig1}a).

As we saw, at the Kondo insulating chemical potential $\mu =0.1$ and for $W=0.35$,
the distribution shows a divergence at $\mathrm{Im}\left[G_{c}^{loc}\left(0-i\delta \right)\right]\approx 10^{-3}$.
A similar diverging \emph{tendency} is observed at $\mu =0$ (Fig.~\ref{cap:fig3}b)
and $W=0.5$. This is precisely the disorder value where the minimum of $\rho _{c}^{typ}$
occurs (cf. Fig.~\ref{cap:fig1}a) and which we have been ascribing to the presence
of many unitary scatterers due to the nearby Kondo insulator. The similarity between
the two distributions strengthens further our case for the importance of the proximity
to the Kondo insulator. Additionally and consistent with this, the divergence is
totally absent at $\mu =-0.5$, where the role played by the Kondo insulator fixed
point is much less important.

It is also interesting to observe in Fig.~\ref{cap:fig3}c, how the Kondo gap is
washed out by disorder: at $W=1.4$, where $\rho _{c}^{typ}$ peaks (Fig.~\ref{cap:fig1}a),
most of the weight of the distribution is already at sizeable values of the DOS and
its shape is very similar to the metallic cases.

\subsection{Distribution of Kondo temperatures}

\label{sub:f-electrons}

We now proceed to the analysis of the physical properties related to the ensemble
of impurity problems. As shown before\cite{mirandavlad1,mirandavlad2} the distribution
of Kondo temperatures of the various f-sites is log-normal for weak disorder, but
broadens and acquires a power-law shape at intermediate values of $W\approx 0.35-0.7$.
Once this power-law becomes singular enough, a Griffiths phase is entered with diverging
thermodynamic responses.\cite{mirandavlad1,mirandavlad2}

\subsubsection{Universality at weak disorder}

We have noticed that for weak disorder, the shape of the distribution of various
quantities, including the Kondo temperature, is universal, irrespective of the shape
of the bare distribution of disorder. %
\begin{figure}
\begin{center}\includegraphics[  width=2.5in,
  keepaspectratio]{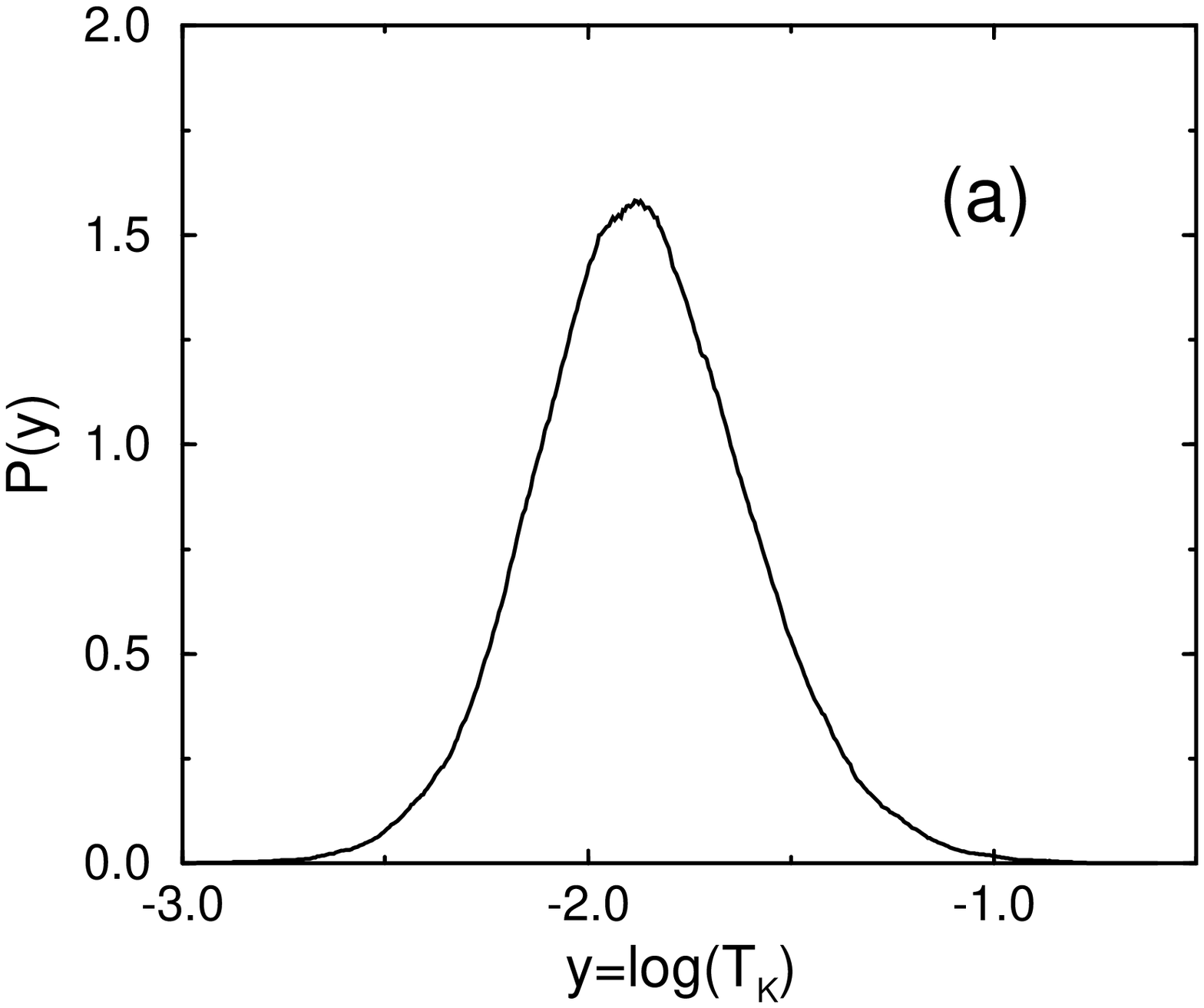}\end{center}

\begin{center}\includegraphics[  width=2.5in,
  keepaspectratio]{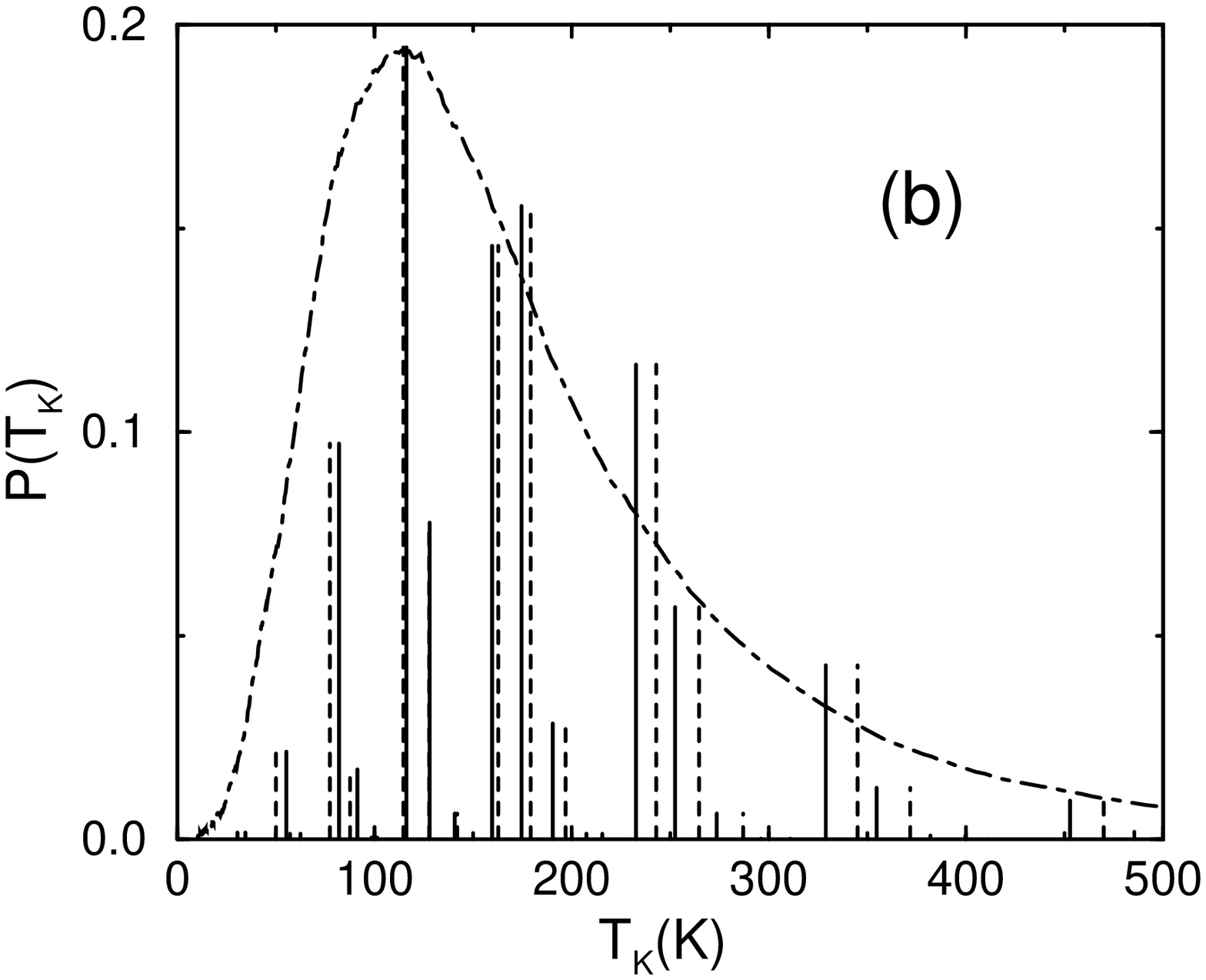}\end{center}

\caption{\label{cap:fig4}(a) Distribution of the logarithm of the Kondo temperature for
a discrete bare distribution of hybridizations, taken from Ref.~\onlinecite{boothetal},
using slave boson mean field theory as the impurity solver (here, we use $t=0.5$,
$E_{f}=-5.54$ and $\mu =-0.2$); (b) Comparison between the smooth distribution
of Kondo temperatures (dot-dashed line) obtained in the statDMFT and the discrete
results of DMFT (vertical solid lines) and the Kondo disorder model (vertical dashed
lines).}
\end{figure}

A nice illustration of this effect is given by the case where the bare
disorder is a \emph{discrete} distribution. As an example, we take the
discrete distribution of hybridization strengths,
$P_{2}\left(V\right)$, determined in Ref.~\onlinecite{boothetal} from
XAFS measurements in UCu$_{5-x}$Pd$_{x}$. The resulting distribution
of $\log T_{K}$ is shown in Fig.~\ref{cap:fig4}a. It is continuous and
has a log-normal shape.  In Fig.~\ref{cap:fig4}b we show the
distribution of $T_{K}$ (smooth dot-dashed line) and compare it to the
discrete distribution obtained in the DMFT (vertical solid lines),
which corresponds to the limit of infinite coordination. We also
include in the figure the results of the Kondo disorder
model\cite{bernaletal,boothetal} (KDM) (vertical dashed lines), which
is very similar to the DMFT. The only differences between the KDM and
the DMFT are that, in the former, no self-consistency is imposed and a
Kondo instead of an Anderson lattice model is used. The difference
between the results of the KDM/DMFT and the statDMFT is striking. The
fluctuations of the conduction electron wave functions incorporated in
the statDMFT smooth out the discrete results of the DMFT into a
universal continuous form. A description of the NFL behavior within
the KDM/DMFT theory would be clearly impossible.  This comparison also
shows that this level of hybridization disorder alone is not able to
generate non-Fermi liquid behavior \emph{ even in statDMFT}, since the
distribution of Kondo temperatures goes to zero as $T_{K}\rightarrow
0$, a point that was stressed in
Refs.~\onlinecite{boothetal,baueretal} However, if disorder in the
conduction electron sites, $P_{1}\left(\epsilon
\right)$, is also included a singular behavior can be
obtained (not shown). The inclusion of conduction electron disorder is
reasonable in UCu$_{5-x}$Pd$_{x}$, since the Cu-Pd interchange affects
both $V_{j}$ and $\epsilon _{j}$.

\subsubsection{Emergence of the Electronic Griffiths phase}

\begin{figure}
\begin{center}\includegraphics[  width=2.5in,
  keepaspectratio]{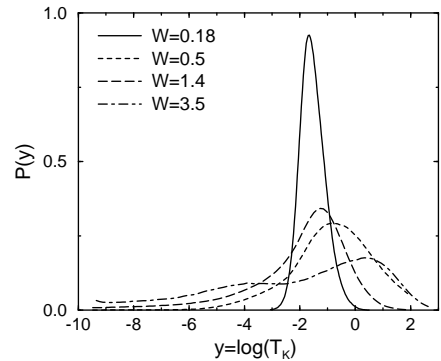}\end{center}

\caption{\label{cap:fig5}Distribution of the logarithm of the Kondo temperature for different
values of disorder, using slave boson mean field theory as the impurity solver ($V=0.5$,
$E_{f}=-1$ and $\mu =0.2$).}
\end{figure}

\begin{figure}
\begin{center}\includegraphics[  width=2.5in,
  keepaspectratio]{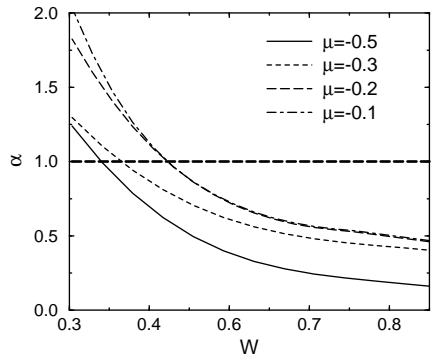}\end{center}

\caption{\label{cap:fig6} Exponent of the power-law distribution of Kondo temperatures as
a function of disorder for different values of the chemical potential ($V=0.5$,
$E_{f}=-1$). The horizontal dashed line indicates the critical value for the emergence
of NFL behavior, where $P\left(T_{K}\right)\propto \mathrm{const.}$ and $\chi \left(T\right)\propto C\left(T\right)/T\propto \ln \left(T_{0}/T\right)$.}
\end{figure}

In order to identify the emergence of the Griffiths phase, we next
study the evolution of the distribution of Kondo temperatures as the
width $W$ of $P_{1}\left(\epsilon \right)$ is varied (with no disorder
in $V$). Typical results are shown in Fig.~\ref{cap:fig5},
corresponding to $\mu =0.2$. As the disorder increases, we find that
the overall width, but most significantly, the size of the low-$T_{K}$
tail rapidly grows. These tails assume a \emph{power-law} form
$P(T_{K})\sim T_{K}^{\alpha -1}$, with the power $\alpha (W)$ being a
monotonically decreasing function of disorder. 
Once again, this behavior cannot be obtained in the rigid scheme of
the KDM/DMFT without unjustified fine tuning.
The thermodynamic response of the system assumes a singular, non-Fermi
liquid form as soon as $\alpha \leq 1$, which happens for sufficiently
strong disorder $W\geq W_{NFL} \approx 0.35-0.45$. Since this behavior
does not reflect any thermodynamic phase transition, it assumes the
character of an electronic Griffiths phase. Here, singular behavior
emerges due to the presence of exponentially rare events (in our case
Kondo spins) which nevertheless provide an exponentially large
contribution to thermodynamic and transport properties and thus
dominate the macroscopic behavior of the system. The $W$-dependence of
the exponent $\alpha $ can be easily obtained by fitting the tails of
these distributions; a representative behavior for several values of
the chemical potential is shown in Fig.~\ref{cap:fig6}.

\subsubsection{NFL and the proximity to the Kondo insulator}

In the previous section we have seen how the proximity of the Kondo insulator plays
a crucial role in determining the disorder dependence of the localization and transport
properties, and leads to the surprising ``bad metal'' behavior for a wide parameter
range. This was argued to reflect the enhanced sensitivity to disorder of those electronic
states that are very close to the Kondo insulator, and which are most easily affected
by localization effects. We have also established that NFL behavior also emerges
as a result of disorder-induced density of states fluctuations. It is then natural
to ask how sensitive this emergence of NFL behavior is to the proximity to the Kondo
insulator, which in the clean limit emerges only in a narrow parameter range, close
(in our case) to $\mu \sim 0.1$.

\begin{figure}
\begin{center}\includegraphics[  width=2.5in,
  keepaspectratio]{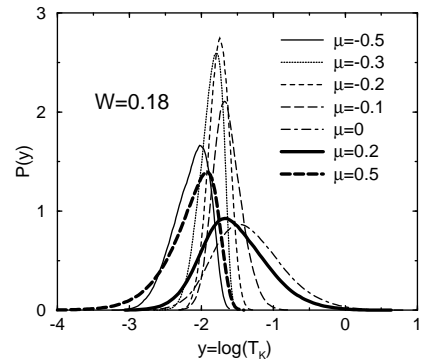}\end{center}

\caption{\label{cap:fig7}Distribution of the logarithm of the Kondo temperature for different
values of the chemical potential and fixed disorder $W=0.18$, using slave boson mean
field theory as the impurity solver ($V=0.5$, $E_{f}=-1$).}
\end{figure}

To address this question, we have systematically investigated the evolution of $P(T_{K})$
as a function of the distance to the Kondo insulator, i.e. as a function of the chemical
potential $\mu $. The behavior for weak disorder ($W=0.18$) is shown in Fig.~\ref{cap:fig7}.
Despite what one would naively expect, these result clearly demonstrate that the
distributions are the broadest \emph{far} from the Kondo insulator. As we can see
on this figure, the distributions narrow down as the Kondo insulator is approached
from either side. As a result, we may expect that the critical disorder strength
$W_{NFL}$ necessary for the emergence of NFL behavior should \emph{increase} closer
to the Kondo insulator. This surprising result is confirmed by examining the $\mu $-dependence
of the exponent $\alpha $ as shown in Fig.~\ref{cap:fig6}. As we can see there,
for a given $W$, the exponent $\alpha $ is indeed \emph{smaller}, and $W_{NFL}$
decreases for larger $\mu $ (far from the Kondo insulator).

At first sight, these findings seem in contradiction to what one may expect, since
we have found that the typical density of states decreases close to the Kondo insulator.
Naively, one could then expect the Kondo temperatures to be depressed as well, leading
to broader distributions and enhanced NFL behavior. On the other hand, we know that
the Kondo temperature remains \emph{finite} within the Kondo insulator, despite the
fact that the density of states at the Fermi energy vanishes there. Although surprising
at first sight, this curious feature of Kondo insulators is at present well understood.
It is called the ``strong coupling Kondo effect''.\cite{vladgabiSCKondo,SCKE1,SCKE2}
It reflects the fact that the Kondo screening is not determined only by electronic
states precisely \emph{at} the Fermi energy, but also by all the states in an energy
interval of order $T_{K}$ around the Fermi energy. Indeed, the average value of
the Kondo temperature (see Fig.~\ref{cap:fig7}) is the \emph{highest} precisely
near the Kondo insulator. In the presence of disorder, the value of the Kondo temperature
is determined by a certain weighted average of the density of states over this extended
energy interval. When localization is present, only the states closest to the Kondo
gap band edge will be appreciably affected, but since not only those states determine
$T_{K}$, the net effect is washed away. We thus conclude that the proximity to the
Kondo insulator, in contrast to transport, does not have appreciable effect on the
emergence of the NFL behavior. Indeed, the critical value of disorder $W_{NFL}$
required for the emergence of NFL behavior is found to have a remarkably weak $\mu $
dependence.

\subsubsection{Universality at strong disorder}

\begin{figure}
\begin{center}\includegraphics[  width=2.5in,
  keepaspectratio]{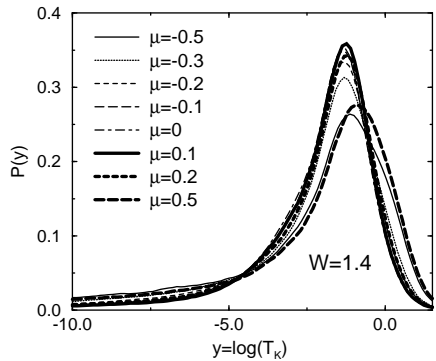}\end{center}

\caption{\label{cap:fig8}Distribution of the logarithm of the Kondo temperature for different
values of the chemical potential and fixed disorder $W=1.4$, using slave boson mean
field theory as the impurity solver ($V=0.5$, $E_{f}=-1$).}
\end{figure}

\begin{figure}
\begin{center}\includegraphics[  width=2.5in,
  keepaspectratio]{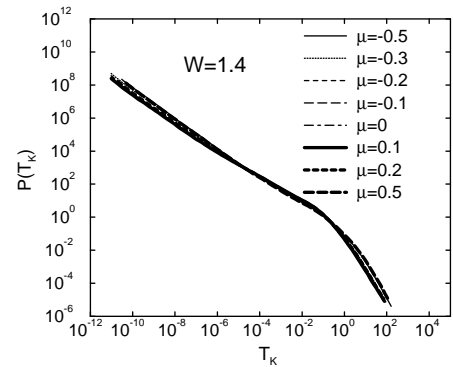}\end{center}

\caption{\label{cap:fig9}Distribution of Kondo temperatures on a log-log scale, for different
values of the chemical potential and fixed disorder $W=1.4$, using slave boson mean
field theory as the impurity solver ($V=0.5$, $E_{f}=-1$).}
\end{figure}

At strong disorder we expect the density of states fluctuations to completely wash
out any trace of the Kondo gap, and in addition to broaden the conduction band, making
it very flat and featureless. As a result, all quantities are expected to have an
extremely weak $\mu $ dependence, leading to a more universal behavior of all quantities.
Such behavior is indeed seen at sufficiently strong disorder, where the typical DOS
curves (Fig.~\ref{cap:fig1}) are seen to merge around $W\sim 1.4$. Similar behavior
is seen in Fig.~\ref{cap:fig8}, which shows $P(\log \, T_{K})$ for different $\mu $'s
at $W=1.4$. This universal behavior is even more striking if this distribution is
plotted on a log-log scale (Fig.~\ref{cap:fig9}), where an almost perfect power-law
tail ($\alpha \sim 0.2$) is seen to span several decades for all the
values of $\mu $.
This is a remarkable example of universality generated by DOS
fluctuations, completely absent in the KDM/DMFT treatment.

%

\subsection{Distribution of the hybridization function}

\begin{figure}
\begin{center}\includegraphics[  width=2.5in,
  keepaspectratio]{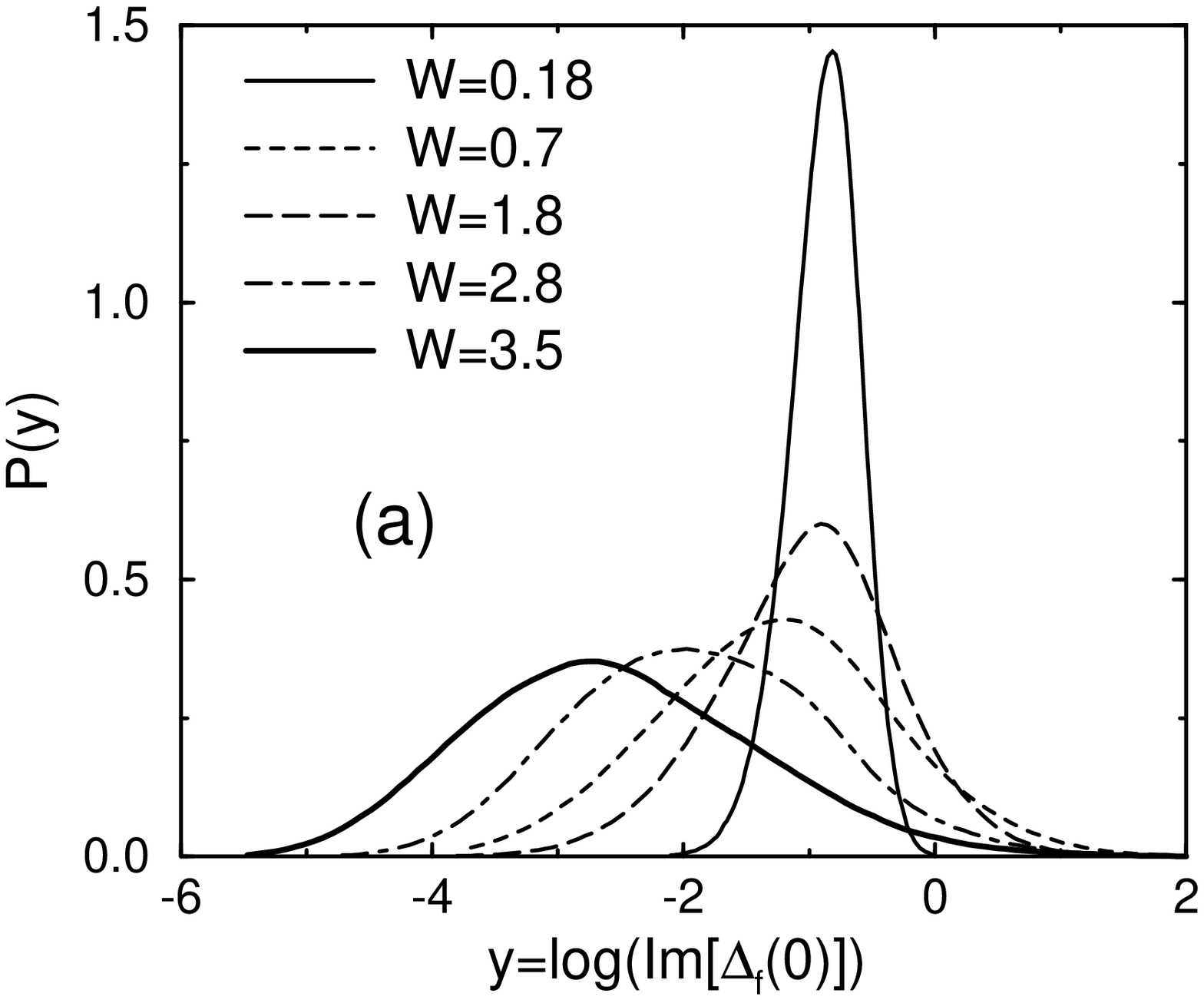}\end{center}

\begin{center}\includegraphics[  width=2.5in,
  keepaspectratio]{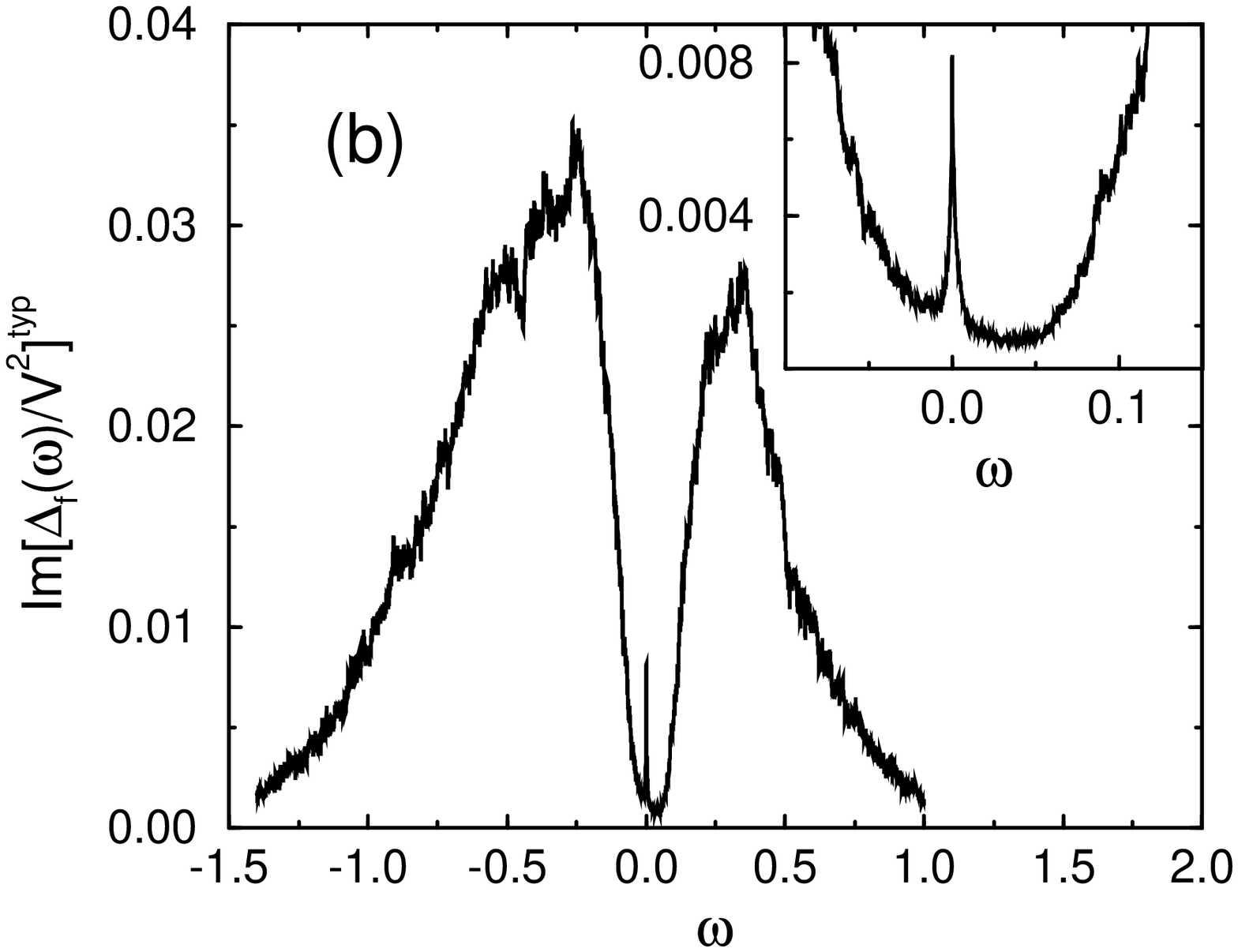}\end{center}

\caption{\label{cap:fig10}(a) Distribution of the logarithm of $\mathrm{Im}\left[\Delta _{f}\left(0\right)\right]$
for different values of disorder and (b) typical value of $\mathrm{Im}\left[\Delta _{f}\left(\omega -i\delta \right)\right]$
as a function of frequency at $W=3.5$, both using slave boson mean field theory as
the impurity solver. The inset in (b) details the behavior close to the chemical
potential ($V=0.5$, $E_{f}=-1$ and $\mu =0.2$).}
\end{figure}

A key input to the determination of the Kondo temperature is the hybridization function
$\Delta _{f}\left(i\omega _{n}\right)$ of Eq.~(\ref{eq:hybfunc}). We show in Fig.~\ref{cap:fig10}a
the distribution of its imaginary part calculated at the chemical potential ($i\omega _{n}\rightarrow 0-i\delta $).
Note that, for a featureless bath, it appears in the exponential of the Kondo temperature
formula, Eq.~(\ref{eq:kondotemp}), which is thus very sensitive to it. It can be
seen in Fig.~\ref{cap:fig10}a that its distribution is very regular and retains
a bell-shaped structure for any $W$. It also inherits the non-monotonic behavior
observed in the conduction electron local DOS (cf. Fig.~\ref{cap:fig1}b).

It is tempting to try to calculate the distribution of Kondo temperatures from the
distribution of $\mathrm{Im}\left[\Delta _{f}\left(0-i\delta \right)\right]$, by
naively applying the Kondo temperature formula, Eq.~(\ref{eq:kondotemp}). This
procedure fails, however: the actual distribution of $T_{K}$ has much lower weight
at small $T_{K}$ values than is predicted by the Kondo temperature formula. The
explanation for this failure lies in the fact, already alluded to before, that $T_{K}$
is determined by a weighted average of $\mathrm{Im}\left[\Delta _{f}\left(\omega -i\delta \right)\right]$
over a region around the Fermi level. This can be glimpsed from the strong frequency
dependence of the \emph{typical} (geometric average) hybridization function close
to the chemical potential, as shown, for frequencies on the real axis, in Fig.~\ref{cap:fig10}b.
It shows a robust and well-defined pseudo-gap, inherited from the nearby Kondo insulator,
and a tiny narrow peak at the chemical potential. This narrow peak is easy to understand:
spatial fluctuations due to disorder give rise to narrow peaks within the pseudo-gap,
most typically at the chemical potential. However, as we have remarked before, the
Kondo temperature can be finite \emph{even if the density of states is zero or almost
zero at the chemical potential}.\cite{vladgabiSCKondo,SCKE1,SCKE2} In this case,
the spectral weight right at the chemical potential is unimportant for the determination
of the Kondo temperature. It is dominated by a whole range of spectral density away
from the Fermi level. Since far from the pseudo-gap region the density of states
is much larger and hence much less affected by the spatial fluctuations, the distribution
of Kondo temperatures is narrower than one might guess based on the distribution
of $\mathrm{Im}\left[\Delta _{f}\left(0-i\delta \right)\right]$ and the Kondo temperature
formula.

When considered together, the results of Section~\ref{sec:Slave-boson} show the
importance of a self-consistent solution of the problem, with a non-trivial interplay
between spatial fluctuations due to localization and strong correlation effects.
However, an important feature that is missed in the slave boson treatment of the
impurity problems is the presence of inelastic scattering. This will be considered
in the next section, where we show the results obtained with perturbation theory
in the interaction.

\section{Perturbation theory results}

\label{sec:Perturbation-theory}

We consider now the results obtained at \textit{finite} $T$ using second order perturbation
theory as the impurity solver.

\subsection{Results for a fixed temperature}

\begin{figure}
\begin{center}\includegraphics[  width=2.5in,
  keepaspectratio]{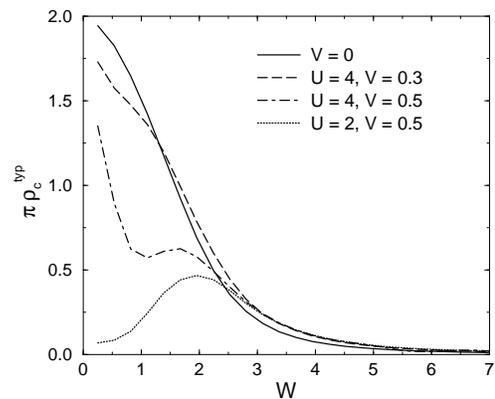}\end{center}

\caption{\label{cap:fig11} Typical density of states near the Fermi surface as a function
of disorder for different values of the hybridization $V$ and the interaction $U$,
using perturbation theory as the impurity solver. Other parameters used were $T=0.003$,
$E_{f}=-1$ and $\mu =0$.}
\end{figure}

\begin{figure}
\begin{center}\includegraphics[  width=2.5in,
  keepaspectratio]{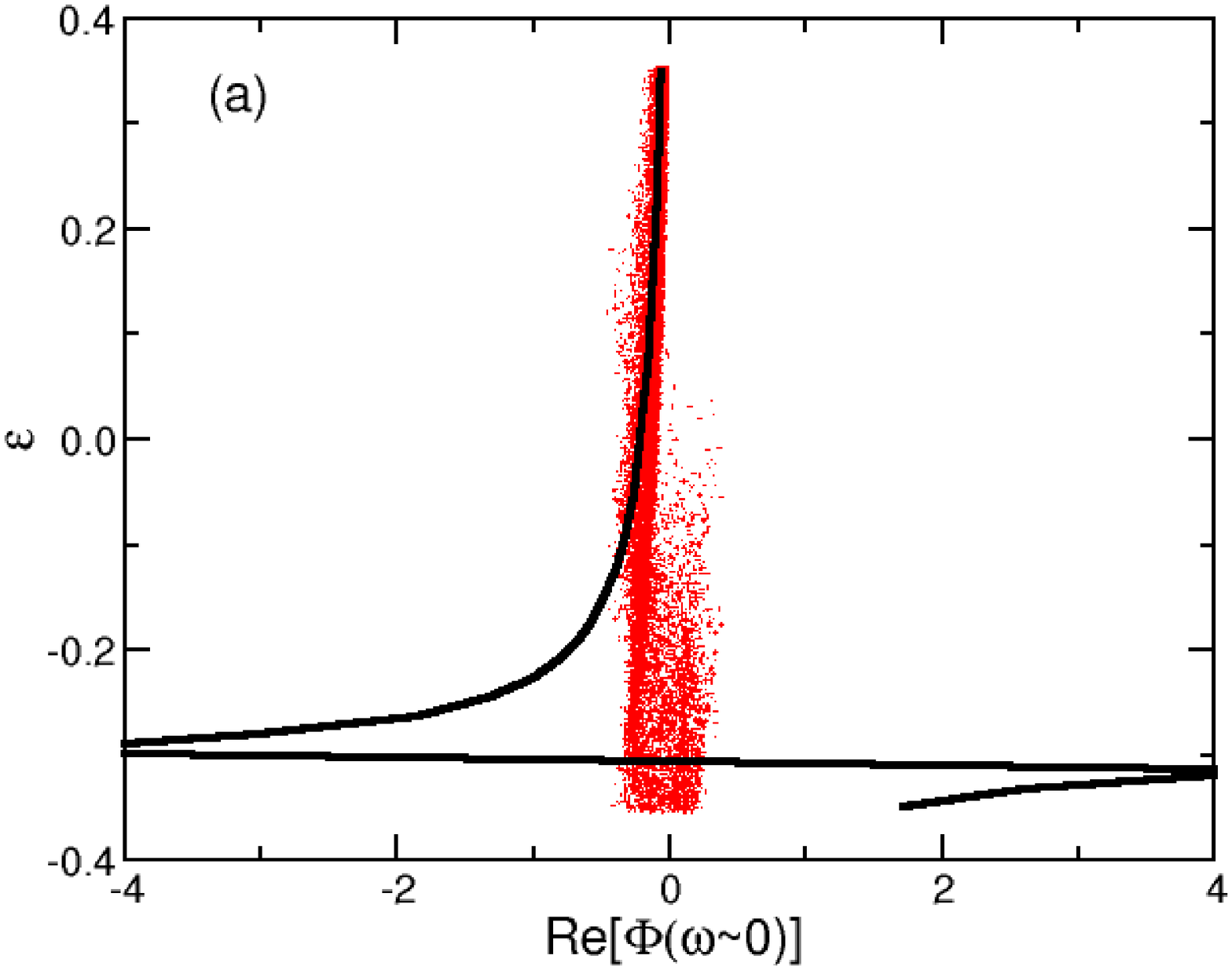}\end{center}

\begin{center}\includegraphics[  width=2.5in,
  keepaspectratio]{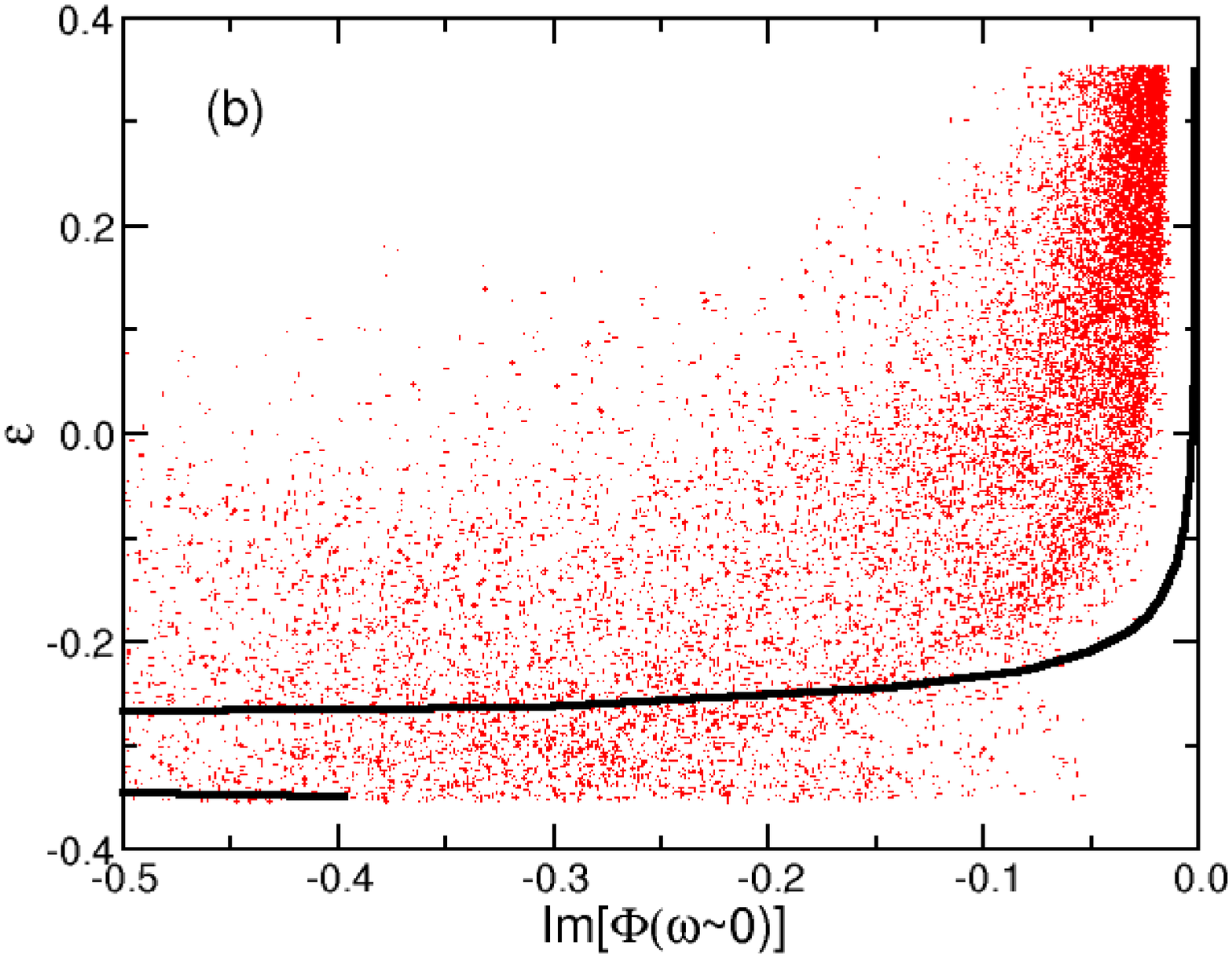}\end{center}

\caption{\label{cap:fig12} Scatter plot of the effective f-shell potential $\Phi _{j}(\omega \approx 0)$
and the bare potential $\epsilon _{j}$ at the same site (dots) for $V=0.3$ and
$W=0.7$, using perturbation theory as the impurity solver: (a) real and (b) imaginary
parts. The solid line is the DMFT result for $W=0.7$. Other parameters used were
$U=4$, $E_{f}=-1$, $T=0.003$ and $\mu =0$.}
\end{figure}

\begin{figure}
\begin{center}\includegraphics[  width=2.5in,
  keepaspectratio]{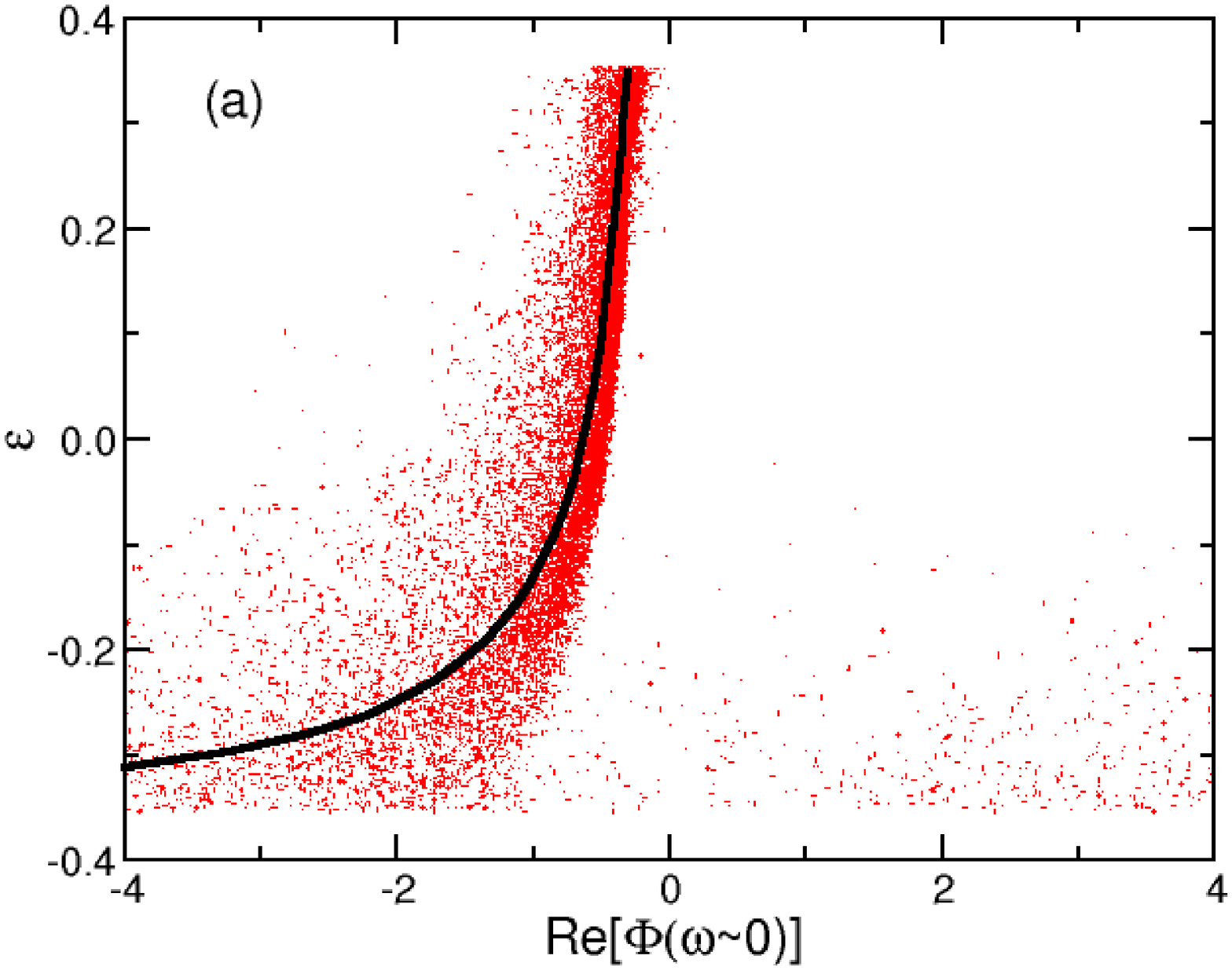}\end{center}

\begin{center}\includegraphics[  width=2.5in,
  keepaspectratio]{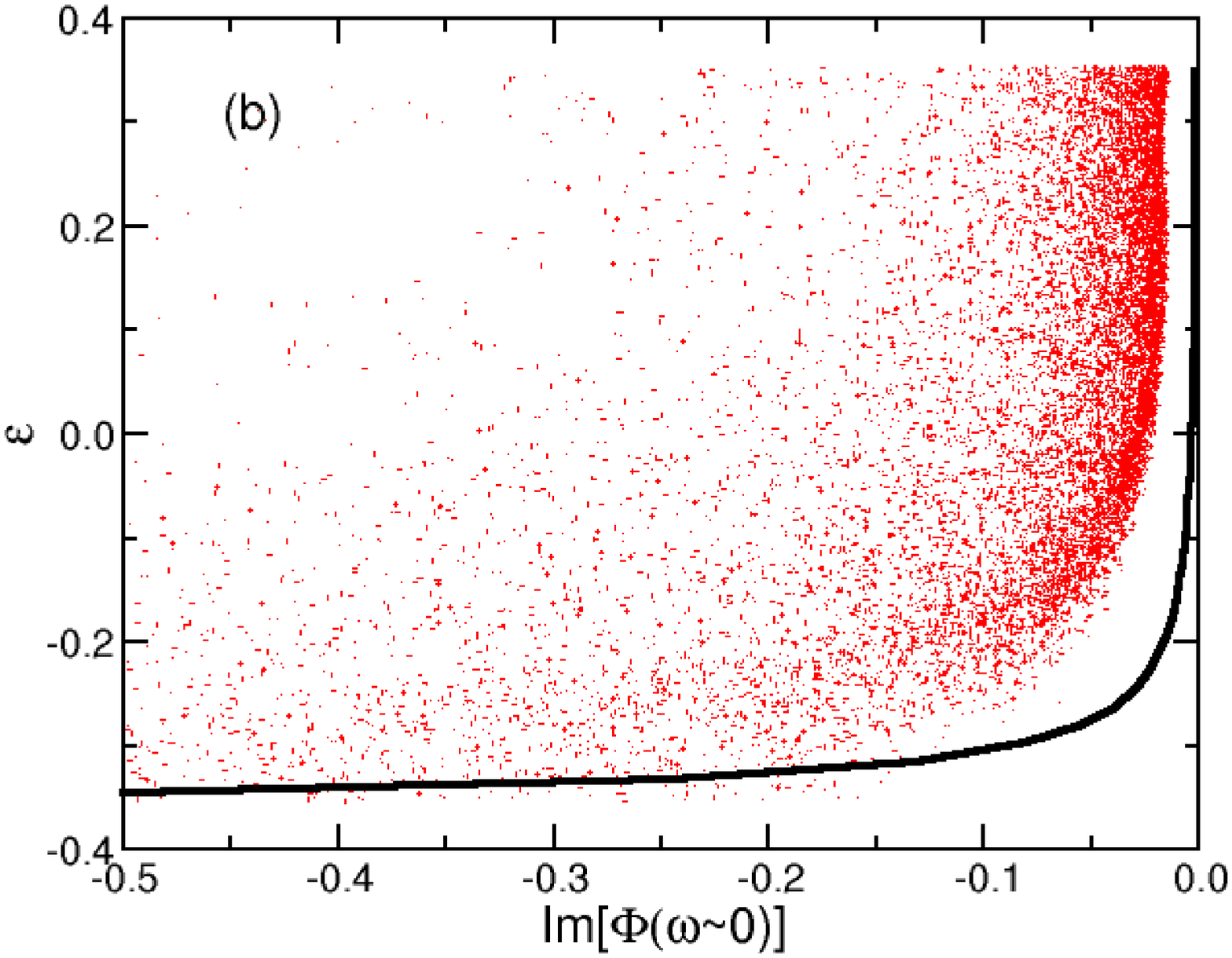}\end{center}

\caption{\label{cap:fig13} Scatter plot of the effective f-shell potential $\Phi _{j}(\omega \approx 0)$
and the bare potential $\epsilon _{j}$ at the same site (dots) for $V=0.5$ and
$W=0.7$, using perturbation theory as the impurity solver: (a) real and (b) imaginary
parts. The solid line is the DMFT result for $W=0.7$. Other parameters as in Fig.~\ref{cap:fig12}.}
\end{figure}

\begin{figure}
\begin{center}\includegraphics[  width=2.5in,
  keepaspectratio]{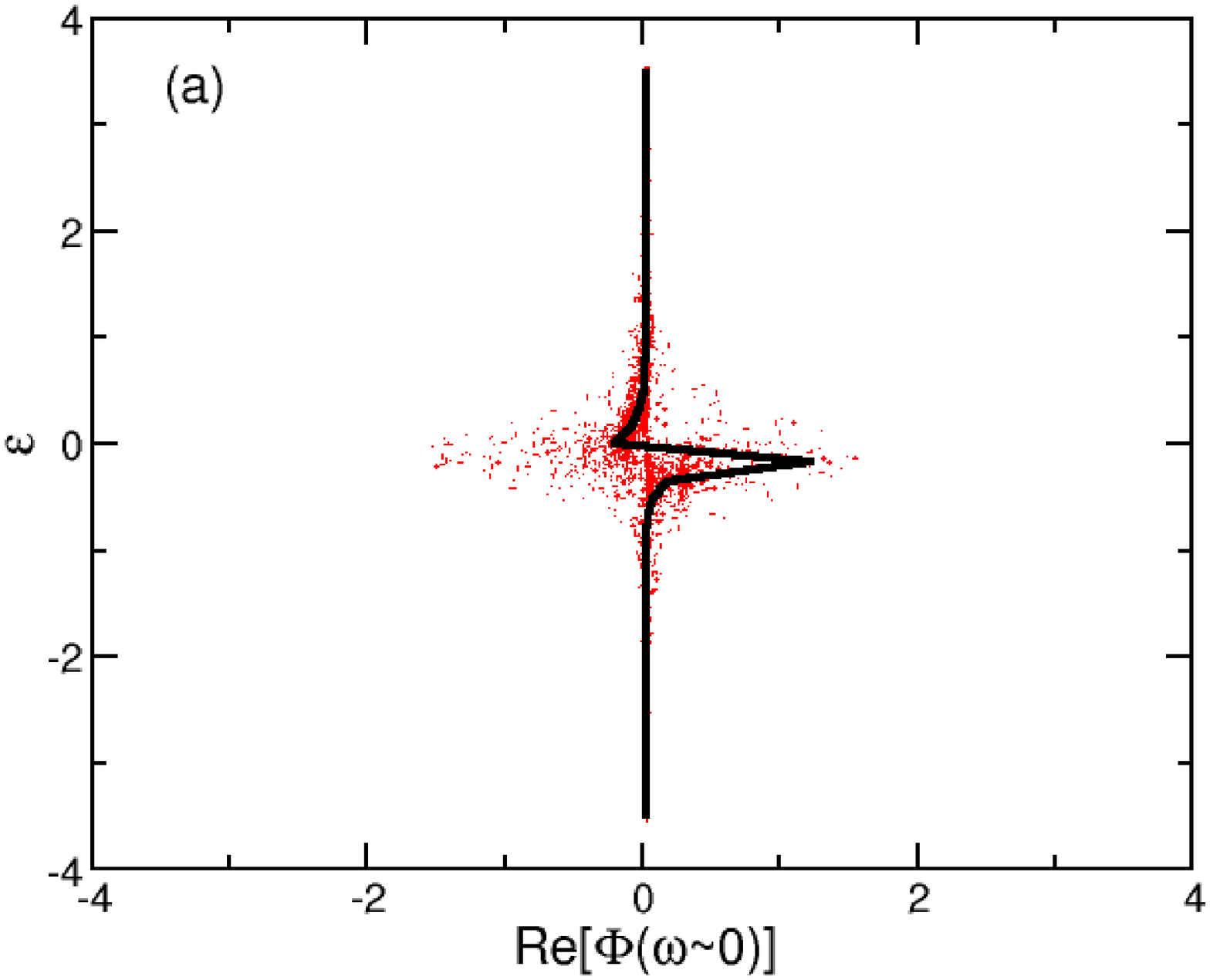}\end{center}

\begin{center}\includegraphics[  width=2.5in,
  keepaspectratio]{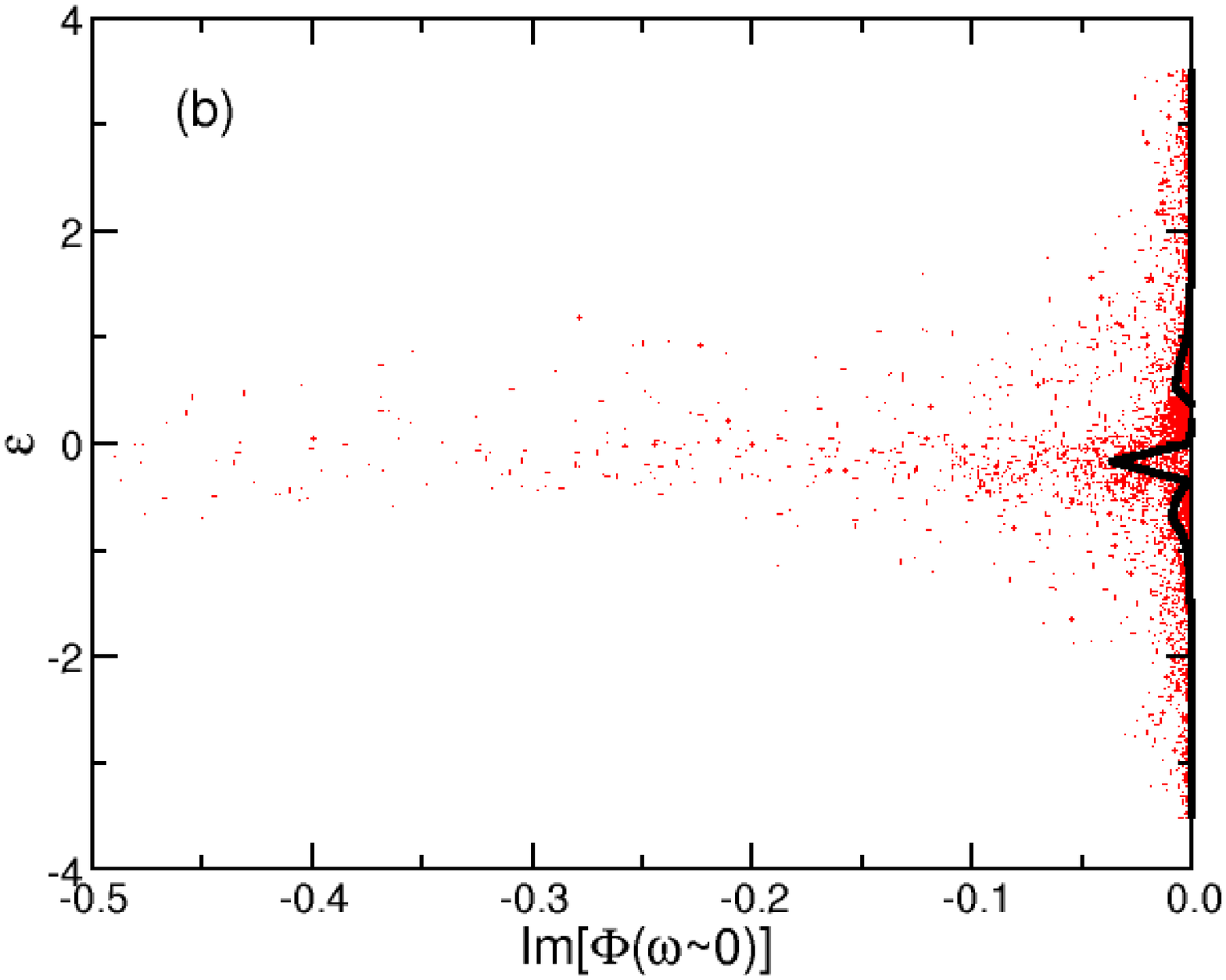}\end{center}

\caption{\label{cap:fig14} Scatter plot of the effective f-shell potential $\Phi _{j}(\omega \approx 0)$
and the bare potential $\epsilon _{j}$ at the same site (dots) for $V=0.3$ and
$W=7$, using perturbation theory as the impurity solver: (a) real and (b) imaginary
parts. The solid line is the DMFT result for $W=7$. Other parameters as in Fig.~\ref{cap:fig12}.}
\end{figure}

\begin{figure}
\begin{center}\includegraphics[  width=2.5in,
  keepaspectratio]{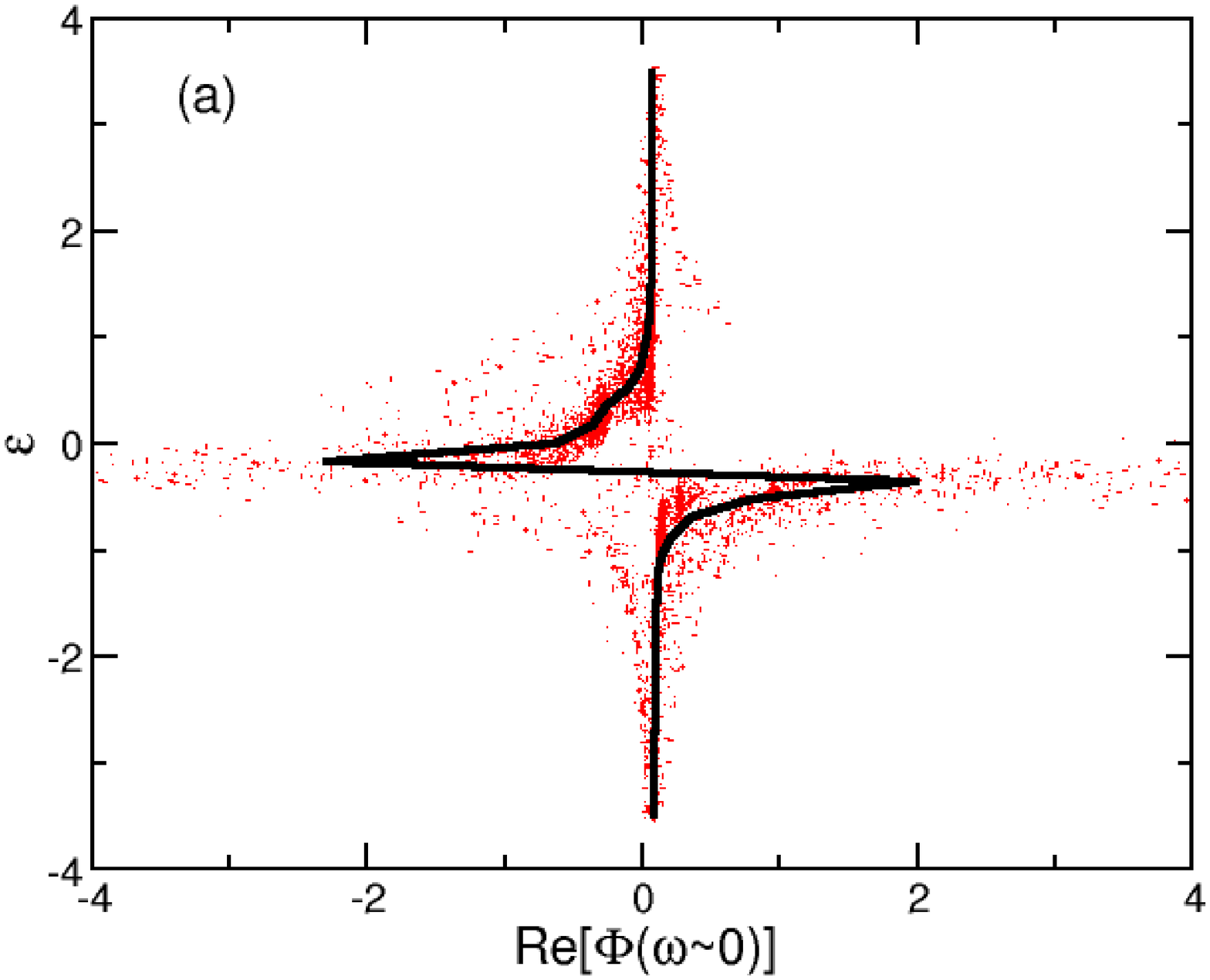}\end{center}

\begin{center}\includegraphics[  width=2.5in,
  keepaspectratio]{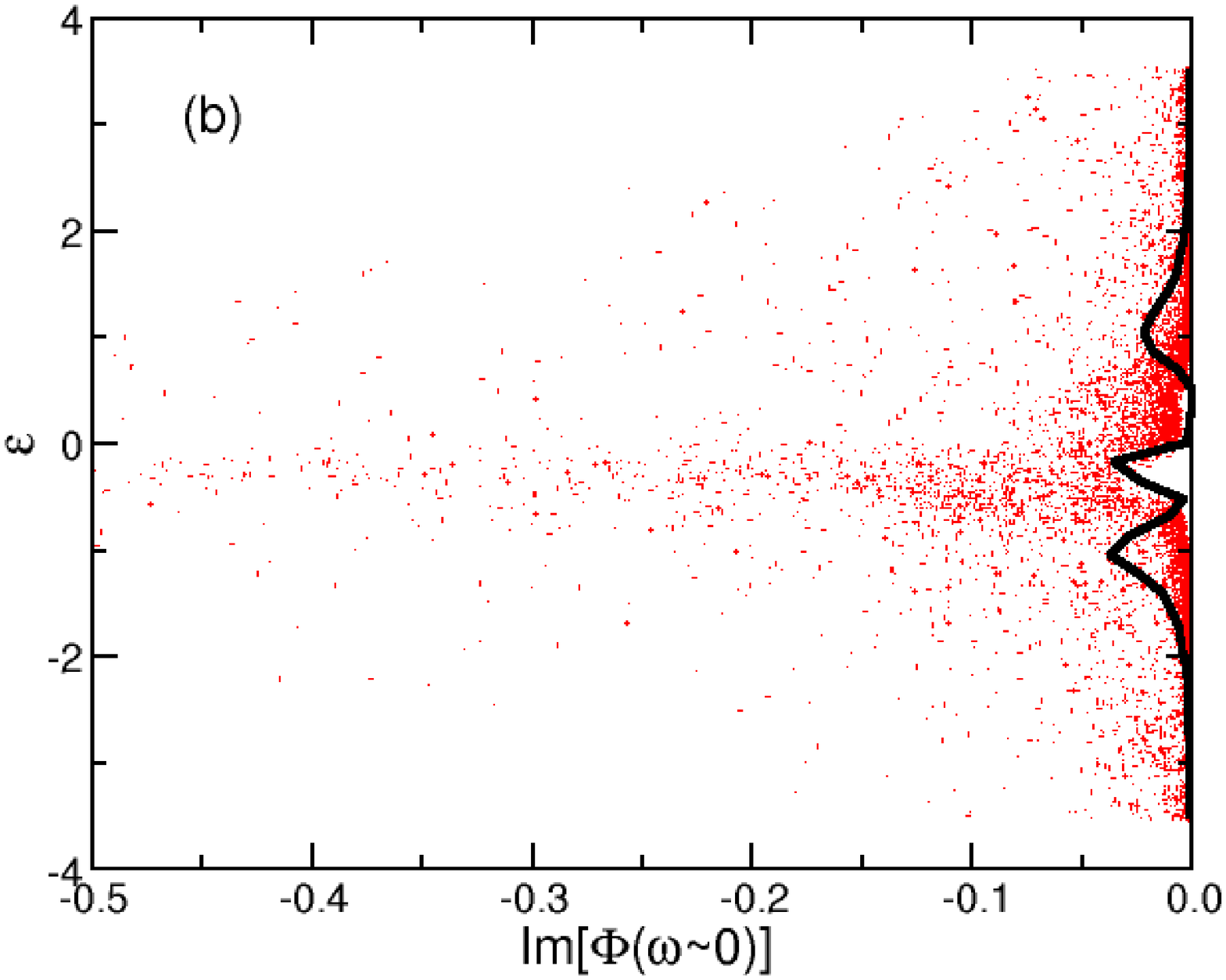}\end{center}

\caption{\label{cap:fig15} Scatter plot of the effective f-shell potential $\Phi _{j}(\omega \approx 0)$
and the bare potential $\epsilon _{j}$ at the same site (dots) for $V=0.5$ and
$W=7$, using perturbation theory as the impurity solver: (a) real and (b) imaginary
parts. The solid line is the DMFT result for $W=7$. Other parameters as in Fig.~\ref{cap:fig12}.}
\end{figure}

Fig.~\ref{cap:fig11} shows the results for the conduction electron typical DOS
near the Fermi level as a function of the disorder parameter $W$ for different values
of the interaction energy $U$ and the hybridization $V$ at $T=0.003$. In order
to understand them, we looked at scatter plots of realizations of the local effective
f-shell potential (real and imaginary parts of $\Phi _{j}(\omega \approx 0)$) and
the corresponding bare potential $\epsilon _{j}$ at the same site, both seen by
the conduction electrons. These results are presented for two different values of
the hybridization, $V=0.3$ and $0.5$, and two different values of the disorder
parameter, $W=0.7$ and $7$, in Figs.~\ref{cap:fig12} to \ref{cap:fig15}. The
other parameters used in the calculation were $U=4$, $T=0.003$, $E_{f}=-1$ and
$\mu =0$. The figures present the results obtained in the statDMFT calculation (represented
by dots) as well as the results of DMFT (full line), in which the disorder treatment
reduces to CPA.\cite{mirandavladgabi1,mirandavladgabi2} It is important to notice
that we have $\mathrm{Im}\left[\Phi (\omega \approx 0)\right]\neq 0$ (as $\mathrm{Im}\left[\Sigma _{f}(\omega \approx 0)\right]\neq 0$
at finite~$T$), implying the presence of \textit{inelastic scattering}, a feature
absent in the slave boson treatment of the last section. Besides, the imaginary part
of the self-energy gets folded into the \emph{real} part of $\Phi \left(\omega \right)$
as well. This is a peculiar feature of a two-band model, where the \emph{effective}
conduction electron self-energy represented by $\Phi \left(\omega \right)$ has a
real part for which inelastic processes contribute. Thus, we should keep in mind
that, even though in an effective description of conduction electron processes $\mathrm{Re}\left[\Phi _{j}(\omega \approx 0)\right]$
and $\epsilon _{j}$ are associated with elastic scattering while $\mathrm{Im}\left[\Phi _{j}(\omega \approx 0)\right]$
is related to inelastic processes, both $\mathrm{Re}\left[\Phi _{j}(\omega \approx 0)\right]$
and $\mathrm{Im}\left[\Phi _{j}(\omega \approx 0)\right]$ contain information on
f-electron collisions.

Let us compare the results for $W=0.7$ for both $V=0.3$ and $V=0.5$, which are in
Figs.~\ref{cap:fig12} and \ref{cap:fig13}. The first observation we make is that
for $V=0.3$ the values of $\mathrm{Re}\left[\Phi (\omega \approx 0)\right]$ are
mainly concentrated around zero, while for $V=0.5$ they are distributed in a wider
range of values. On the other hand, the results for $\mathrm{Im}\left[\Phi (\omega \approx 0)\right]$
show that the inelastic scattering is stronger for $V=0.3$ than for $V=0.5$. Concerning
the results for $V=0.5$, the great concentration of sites with large $\mathrm{Re}\left[\Phi (\omega \approx 0)\right]$
explains the great decrease in the typical DOS for low disorder shown in Fig.~\ref{cap:fig11}.
These sites act as almost unitary scatterers, which give rise to a maximally allowed
scattering phase shift ($\delta =\pi /2$) for the conduction electrons, and represent
droplets of Kondo insulator within the metal, in close parallel to the slave boson
results discussed of Section \ref{sec:Slave-boson}.\cite{mirandavlad1,mirandavlad2}
For $V=0.3$, although the inelastic scattering is stronger, it cannot compensate
for the much narrower distribution of $\mathrm{Re}\left[\Phi (\omega \approx 0)\right]$.
Thus, the DOS does not decrease as fast as for $V=0.5$.

As the disorder increases to intermediate values ($W\approx 1.8$), the distribution
of $\epsilon $ becomes larger, causing a steady decrease in the typical DOS for
$V=0.3$. However, for $V=0.5$, the typical DOS presents a non-monotonic behavior
similar to the slave boson results for $-0.3\alt \mu \alt 0.1$ and $0.1\alt \mu \alt 0.4$.
As we explained in Section \ref{sec:Slave-boson}, this is a result of the fact that,
as the disorder increases, the concentration of unitary scatterers first increases,
then saturates and the bare disorder dominates over the f-related one ($\Phi _{j}\left(\omega \right)$).\cite{mirandavlad1,mirandavlad2}
Indeed, even for $V=0.3$ we notice the presence of a slight {}``shoulder'' in
the typical DOS around $W=1.1$, whose origin is the same as that of the non-monotonic
behavior at $V=0.5$. We call attention to the slave boson results of Fig.~\ref{cap:fig2},
which similarly show that a decrease of the hybridization strength moves the {}``dip-hump''
feature to smaller disorder values. Moreover, as the current calculation was done
at finite temperature, inelastic scattering also plays some role in smoothing out
this feature.

As the disorder continues to increase and the distribution of $\epsilon $ becomes
broader, the typical DOS for both $V=0.3$ and $V=0.5$ decreases. These results
tend to the non-interacting one, as is expected if only the bare disorder plays a
role. Indeed, comparing the results for $W=7$ (Figs.~\ref{cap:fig14} and \ref{cap:fig15})
we notice that the realizations for which the bare disorder $\epsilon $ is large
have $\mathrm{Re}\left[\Phi (\omega \approx 0)\right]$ around zero, meaning that
the real part of the f-shell disorder is not important. Besides, the values of $\mathrm{Im}\left[\Phi (\omega \approx 0)\right]$
are small for these realizations. In the case of the realizations for which $\epsilon $
is small, $\mathrm{Re}\left[\Phi (\omega \approx 0)\right]$ attains values that
are larger than the results for $W=0.7$ (cf. Figs.~\ref{cap:fig12} and \ref{cap:fig13}).

The above discussion has focused on the conduction electron viewpoint. Let us now
consider how the presence of disorder in $\epsilon $ is seen by the f-electrons.
For the auxiliary one-impurity problem, the important scale is the Kondo temperature
$T_{K}$, which measures the coupling between the impurity and the conduction electron
bath. The presence of disorder in $\epsilon $ generates a distribution of $T_{K}$,
as we have seen before. Thus at finite temperature some of the sites have $T_{K}>T$,
forming a singlet state with the bath, while the sites for which $T_{K}<T$ represent
an almost free spin, scattering conduction electrons incoherently. This incoherence
is characterized by a large amount of inelastic scattering, signaled by a significant
imaginary part of the self-energy. Going back to the results for $W=0.7$, the fact
that the inelastic scattering is stronger for $V=0.3$ than for $V=0.5$ reflects
the larger number of incoherent sites (with $T_{K}<T$) in the former, since the
smaller the hybridization, the smaller the Kondo temperature (see Eqs.~\ref{eq:kondotemp}
and \ref{eq:kondoj}). On the other hand, the sites with large $\mathrm{Re}\left[\Phi (\omega \approx 0)\right]$
for $V=0.5$, which are responsible for the great decrease in the typical DOS, represent
sites with $T_{K}>T$.

Fig.~\ref{cap:fig11} also shows the results for the typical DOS for $U=2$, and
$V=0.5$. In this case, the system has particle-hole symmetry in the clean limit,
presenting a gap in its DOS (the Kondo insulator). This explains the fact that for
small disorder the typical DOS decreases as $W$ decreases. The overall behavior
here bears strong similarity with the slave boson results of Fig.~\ref{cap:fig1}a
(circles) and the explanation for it has been given in Section \ref{sec:Slave-boson}.

\subsection{Temperature dependence}

\begin{figure}
\begin{center}\includegraphics[  width=2.5in,
  keepaspectratio]{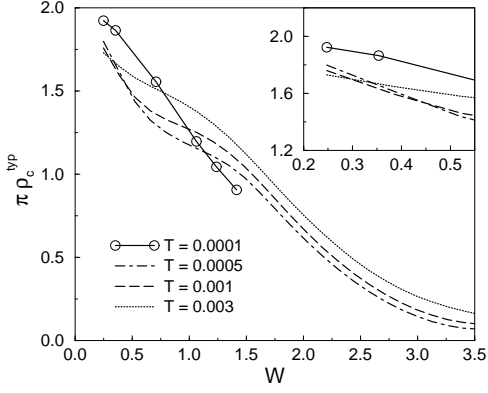}\end{center}

\caption{\label{cap:fig16} Typical density of states near the Fermi surface as a function
of disorder for different values of temperature, using perturbation theory as the
impurity solver. Other parameters used were $U=4$, $V=0.3$, $E_{f}=-1$ and $\mu =0$.}
\end{figure}

\begin{figure}
\begin{center}\includegraphics[  width=2.5in,
  keepaspectratio]{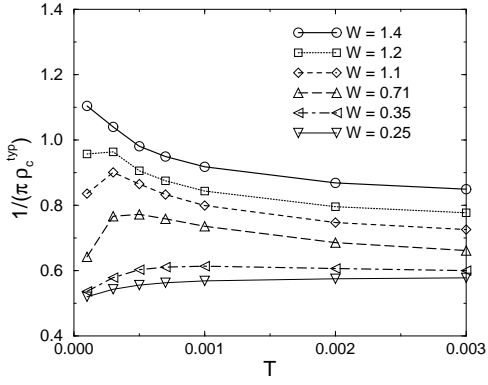}\end{center}

\caption{\label{cap:fig17} Inverse typical density of states near the
Fermi surface as a function of temperature for different values of
disorder, using perturbation theory as the impurity solver. Other
parameters as in Fig.~\ref{cap:fig16}.}
\end{figure}

Fig.~\ref{cap:fig16} presents the results for the typical DOS as a function of
$W$ for different temperatures. The other parameters used were $U=4$, $V=0.3$,
$E_{f}=-1$ and $\mu =0$. Here is where the interplay between inelastic and elastic
processes proves to be fairly non-trivial and the use of a technique that incorporates
both, such as perturbation theory, is crucial. First we note that for the lowest
disorder value ($W=0.25$), the typical DOS decreases with increasing temperature.
This is made more clear in the inset. On the other hand, for $W=1.4$, this tendency
is reversed. Finally, in between these two extremes, the temperature dependence can
be non-monotonic. A better sense of the overall behavior can be grasped by plotting
the inverse of the typical DOS as a function of $T$ for different values of $W$,
as shown in Fig.~\ref{cap:fig17}. It is clear that for $0.7\leq W\leq 1.2$, the
inverse typical DOS shows a peak as a function of $T$, which gradually moves to
zero temperature as disorder is increased.

\begin{figure}
\begin{center}\includegraphics[  width=2.5in,
  keepaspectratio]{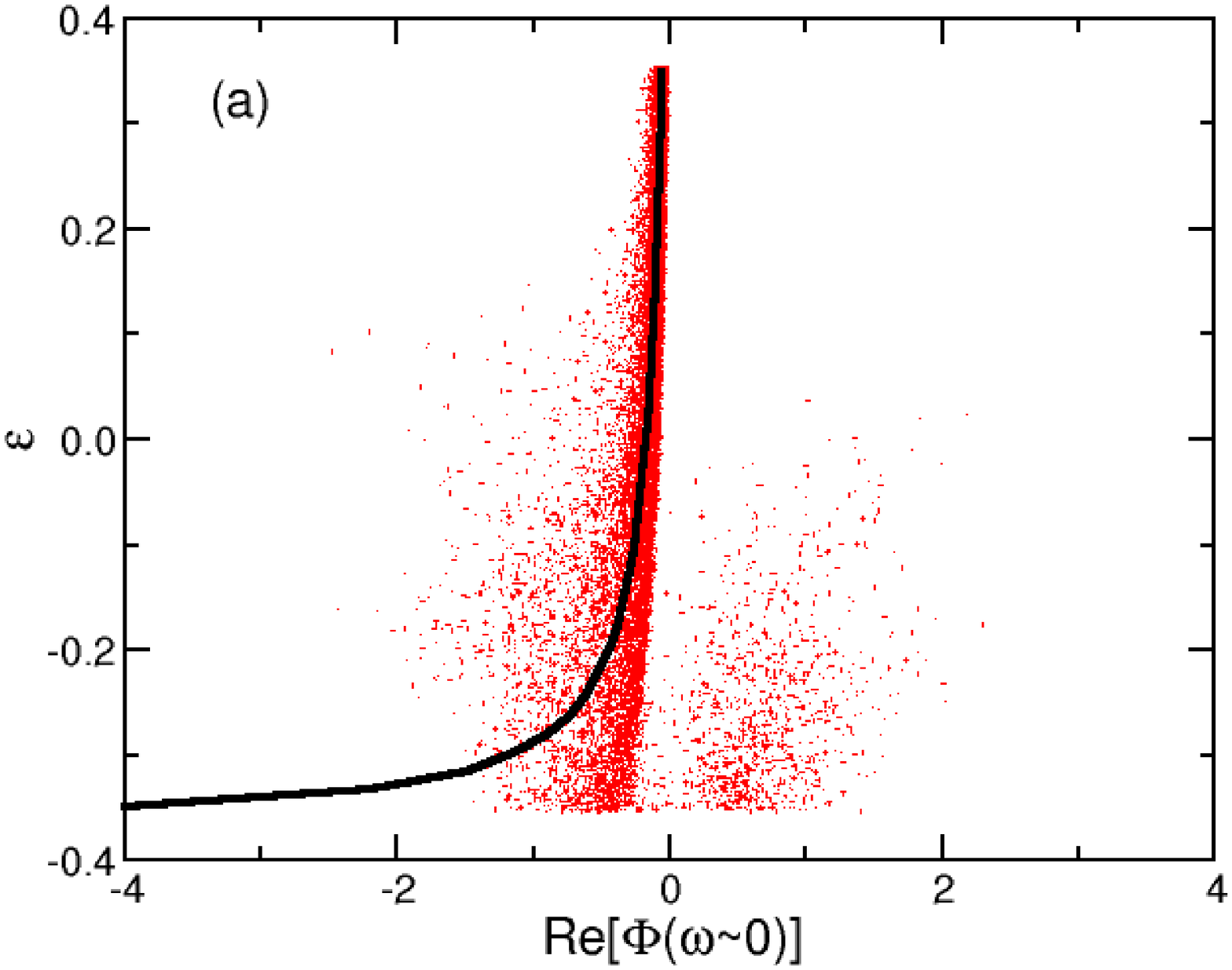}\end{center}

\begin{center}\includegraphics[  width=2.5in,
  keepaspectratio]{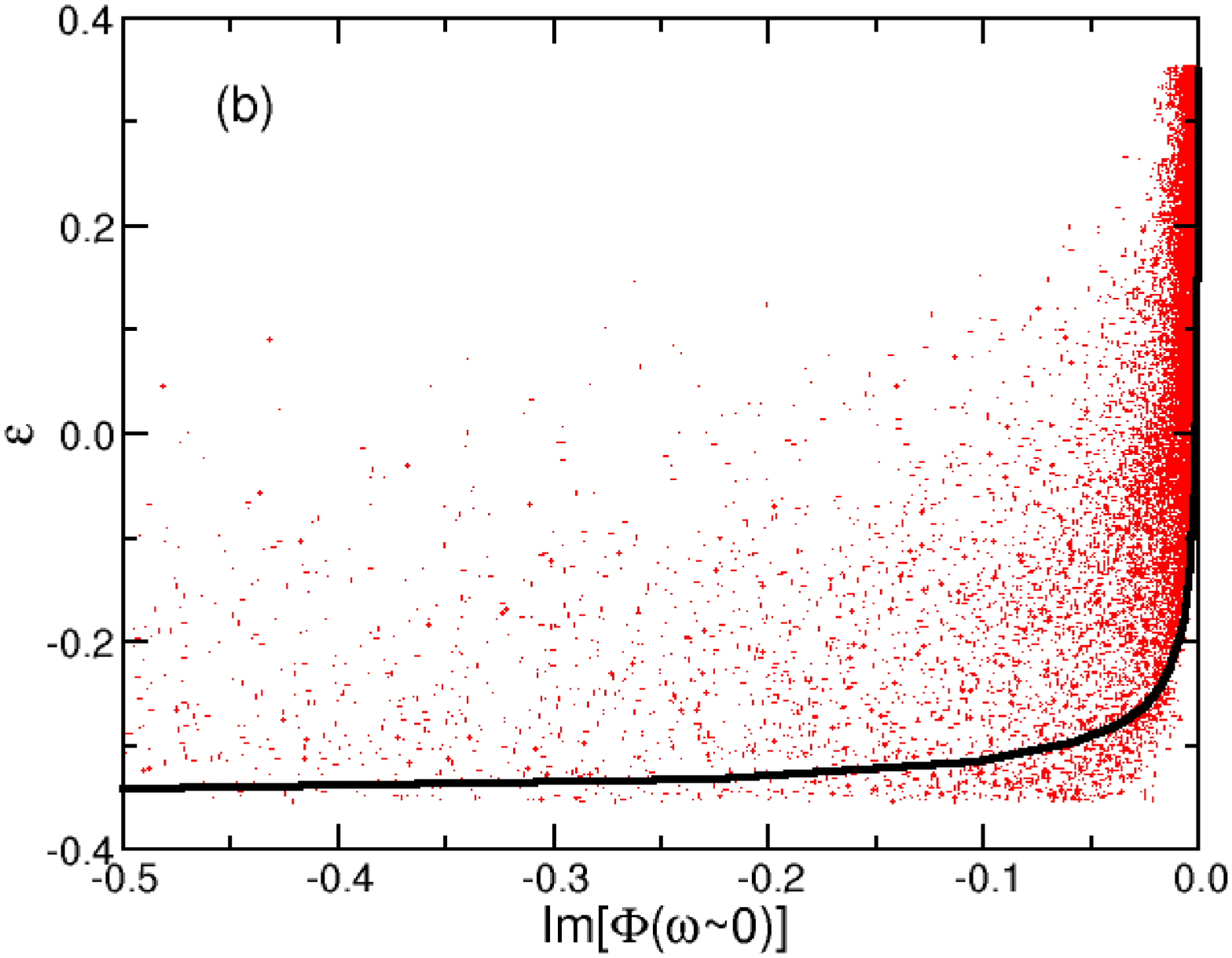}\end{center}

\caption{\label{cap:fig18} Scatter plot of the effective f-shell potential $\Phi _{j}(\omega \approx 0)$
and the bare potential $\epsilon _{j}$ at the same site (dots) for $T=0.0005$ and
$W=0.7$, using perturbation theory as the impurity solver: (a) real and (b) imaginary
parts. The solid line is the DMFT result for $W=0.7$. Other parameters as in Fig.~\ref{cap:fig16}.}
\end{figure}

\begin{figure}
\begin{center}\includegraphics[  width=2.5in,
  keepaspectratio]{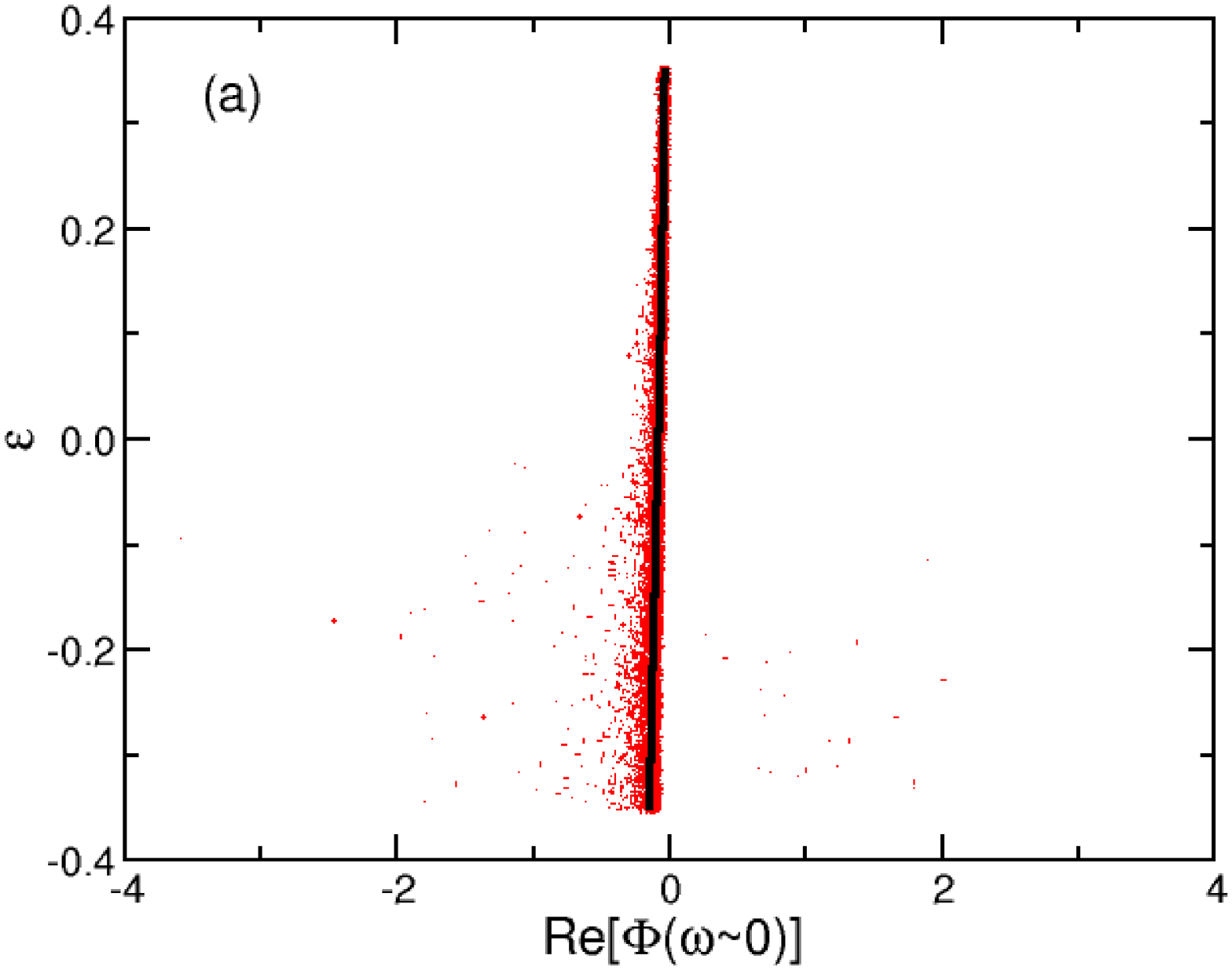}\end{center}

\begin{center}\includegraphics[  width=2.5in,
  keepaspectratio]{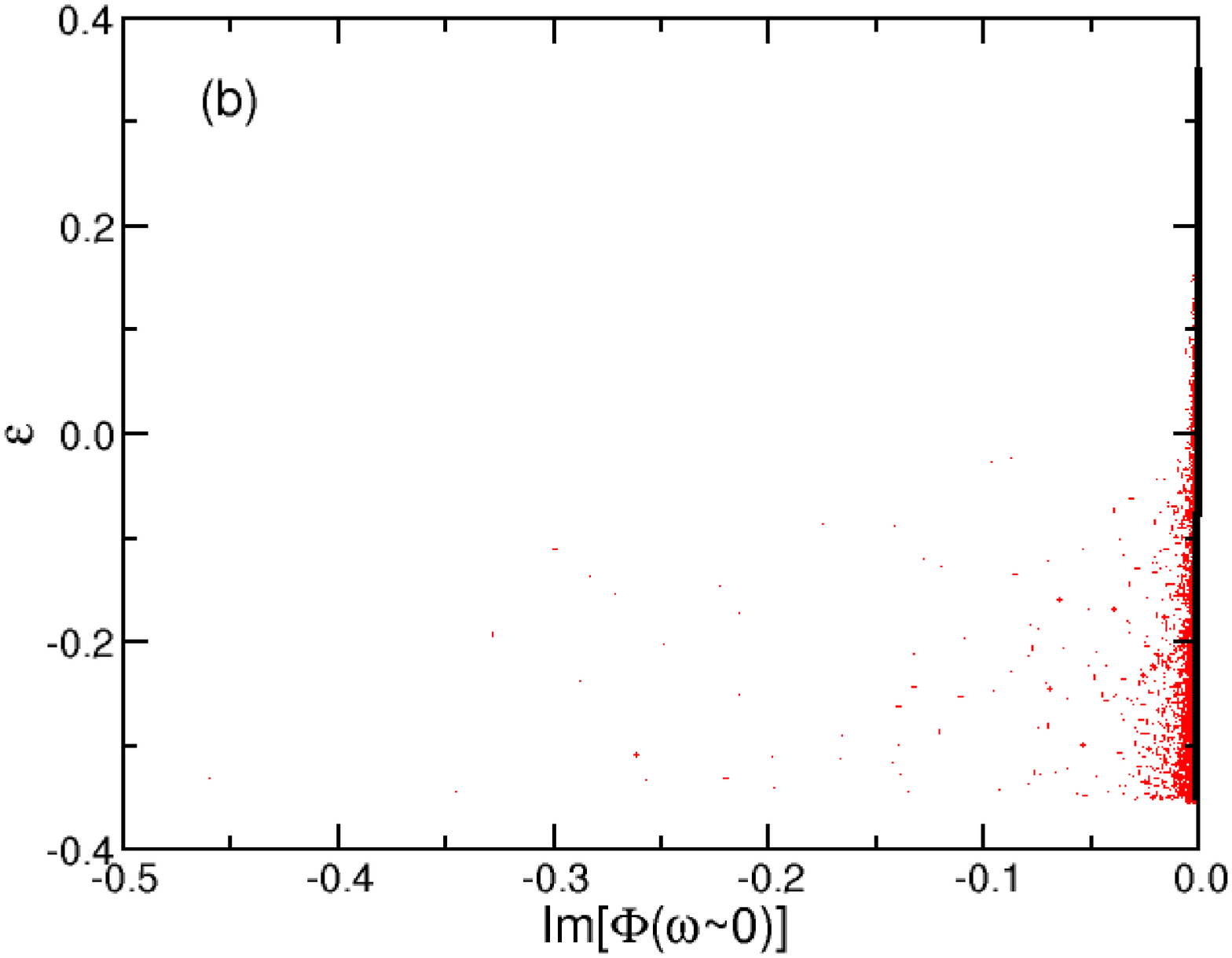}\end{center}

\caption{\label{cap:fig19} Scatter plot of the effective f-shell potential $\Phi _{j}(\omega \approx 0)$
and the bare potential $\epsilon _{j}$ at the same site (dots) for $T=0.0001$ and
$W=0.7$, using perturbation theory as the impurity solver: (a) real and (b) imaginary
parts. The solid line is the DMFT result for $W=0.7$. Other parameters as in Fig.~\ref{cap:fig16}.}
\end{figure}

The temperature dependence of the typical DOS can be rationalized by looking at the
corresponding changes in the distribution of $\Phi \left(\omega \approx 0\right)$.
For this purpose, we focus on the disorder value of $W=0.7$, for which $1/\pi \rho _{c}^{typ}$
has a clear maximum at $T\approx 0.0005$. We show in Figs.~\ref{cap:fig18} and
\ref{cap:fig19} the scatter plots of $\Phi \left(\omega \approx 0\right)$ at $T=0.0001$
and $T=0.0005$, respectively, which should be compared to the higher temperature
results ($T=0.003$) of Fig.~\ref{cap:fig12}. From Fig.~\ref{cap:fig17} we see
that the inverse typical DOS increases as we go from $T=0.0001$ to $T=0.0005$ and
then decreases as the temperature is varied up to $T=0.003$. As can be seen from
the figures, this non-monotonic behavior is governed by the effective f-disorder
encoded in the distribution of $\mathrm{Re}\left[\Phi \left(\omega \approx 0\right)\right]$.
Indeed, its variance increases from $T=0.0001$ to $T=0.0005$ but decreases from
$T=0.0005$ to $T=0.003$. Note that the imaginary part of $\Phi \left(\omega \approx 0\right)$
always increases with increasing temperature, reflecting the enhancement of inelastic
processes. In terms of $T_{K}$, this is the same as saying that as the temperature
increases the number of sites with $T_{K}<T$, which have a stronger inelastic scattering,
becomes larger. However, in the interval from about $T=0.0005$ up to $T=0.003$,
the increase of $\mathrm{Im}\left[\Phi \left(\omega \approx 0\right)\right]$ is
\emph{outweighed} by the narrowing of the distribution of $\mathrm{Re}\left[\Phi \left(\omega \approx 0\right)\right]$,
which is the dominant contribution. The above analysis can be similarly extended
to the other values of disorder shown in Fig.~\ref{cap:fig17}.

All along we have been using the typical DOS as a measure of the conducting properties
of the system. This is justified by its interpretation as a escape rate from a lattice
site and the fact that it vanishes at the localization transition.\cite{andersonloc}
Ideally, one would like to calculate the conductivity instead. This is a difficult
task in the present scheme, however, although an approximate calculation can be performed,
which becomes accurate close to the localization transition.\cite{girvinjonson,tmt}
It requires the calculation of the propagator between two different sites, which
goes beyond our current method, whose focus is on local Green's functions only. Even
in view of all these caveats, however, it is tempting to use the inverse typical
DOS as an approximate measure of the resistivity, \emph{specially far from the weakly
disordered region} ($W>0.35$). If this is done, then the fanning out of the {}``resistivity''
curves of Fig.~\ref{cap:fig17} as $T\rightarrow 0$ is reminiscent of the Mooij
correlations,\cite{mooij} originally observed in disordered transition metal alloys,
but which are also seen in heavy fermion alloys,\cite{linschlottmann} doped Kondo
insulators\cite{ditusaKIprl,ditusaKIprlerr,ditusaKI} and even in two-dimensional
systems as in the metal-oxide-semiconductor field-effect transistors.\cite{kravchenkormp}
Our results and the discussion above show that the interplay between localization
effects and electron-electron interactions can give rise to the Mooij correlations,
without the need to invoke other sources, such as electron-phonon interactions.\cite{girvinjonson,leeramakrishnan}
In addition, we have pointed out how the rapid drop in the typical DOS is a consequence
of the proximity to the Kondo insulator, a region where localization effects are
particularly large. Taken together, these observations point to a close connection
between Mooij correlations and localization effects in the vicinity of a Kondo or
a Mott insulator. It would be interesting to test these ideas by a direct calculation
of the resistivity within a scheme capable of incorporating both localization and
strong correlations.

Finally, regarding the DMFT calculation presented in all the figures discussed in
this Section, we note that the results obtained for $\mathrm{Re}\left[\Phi (\omega \approx 0)\right]$
fall approximately in the region where the concentration of realizations in the statDMFT
calculation is largest. There is no reason to expect DMFT and the statDMFT to give
similar results, except at weak disorder. Nevertheless, our results show that, surprisingly,
DMFT can serve as a rough guide for the most probable values of $\mathrm{Re}\left[\Phi (\omega \approx 0)\right]$.
There is a sizeable discrepancy, however, between the DMFT and the statDMFT results
for $\mathrm{Im}\left[\Phi \left(\omega \approx 0\right)\right]$ at the highest
temperature ($T=0.003$) and low disorder. The DMFT line in this case is an underestimate
of the realizations obtained within the statDMFT. As the temperature is lowered a
better agreement is obtained. Thus, besides its inherent neglect of localization
effects, DMFT should be used only as a lower bound when gauging the importance of
inelastic processes in disordered Anderson lattices.

\section{Discussion and conclusions}

\label{sec:Discussion-and-conclusions}

We have in this paper extensively characterized the physics of
disordered Anderson lattices within the statDMFT scheme, which is able
to incorporate both localization effects and the local correlations
coming from electron-electron interactions. This was done using both
large-N methods and perturbation theory for the auxiliary single
impurity problems. These are in a sense complementary approaches. On
the one hand, large-N theory is ideal for ground state properties,
where inelastic effects are absent. In particular, it affords a quick
and reliable way of calculating Kondo temperatures (with the correct
exponential dependence) and scattering phase shifts at $T=0$.  Without
$1/N$ corrections, however, it is unsuitable for a finite temperature
calculation.  On the other hand, second order perturbation theory has
the advantage of being equally flexible with the added bonus of
incorporating inelastic processes and the temperature dependence of
the scattering phase shifts. Nevertheless, it fails to capture the
exponential nature of the low temperature scale of the single impurity
problem. Taken together, the two methods have enabled us to put on
firm grounds the conclusions laid out in previous
work,\cite{mirandavlad1,mirandavlad2} namely, the emergence of an
electronic Griffiths phase in Anderson lattices governed by the
proximity to the disorder-induced localization transition. In
particular, several inadequacies of the early Kondo Disorder Model
(and its formulation as a Dynamical Mean Field Theory) have been given
a better theoretical basis once localization effects were
included. The self-averaging effect introduced by the spatial
fluctuations of the conduction electron wave functions induce a much
higher degree of universality than is possible in the rigid KDM
description.  Furthermore, the perturbation theory treatment has also
suggested a mechanism behind the ubiquitous observation of Mooij
correlations in the resistivity of disordered materials.

Even within the confines of the approximations of the statDMFT scheme,
there are still outstanding issues that we would like to resolve. Even
though a fully analytical treatment is probably impossible, it might
be feasible to devise an approximate parameterization of the statDMFT
on the Bethe lattice, specially at weak disorder. In this respect, the
universality of the distributions of {}``dressed'' quantities, such as
the various local Green's functions and the Kondo temperatures, which
are either Gaussian or log-normal is a useful guide. A {}``toy'' model
that assumes a simple form for the distribution of $\Delta
_{cj}\left(i\omega _{n}\right)$ can be written down, which recovers
the Griffiths singularities obtained in the numerical
treatment.\cite{vladmirandaunp} This {}``toy'' model may prove useful
for a calculation of the resistivity and for generalizations of the
statDMFT treatment.

On the other hand, a specially important effect neglected in the
statDMFT is two-particle inter-site correlations, particularly in the
spin channel. Indeed, the proliferation of poorly quenched low-$T_{K}$
spins in our treatment generates a large amount of entropy that must
be relieved at low temperatures through inter-site correlations.
Indeed, experimental evidence in favor of spin-glass dynamics at low
temperatures in UCu$_{5-x}$Pd$_{x}$\cite{dougetal} and
URh$_{2}$Ge$_{2}$\cite{sullowetal} makes the inclusion of inter-site
correlations more pressing. A promising avenue of attack would be to
remain true to the spirit of the DMFT and use its extended
version. Several treatments along these lines have been
attempted.\cite{senguptageorges,grempelmarcelo1,burdinetal} The
challenge in our case is to incorporate both the dynamical inter-site
correlations of the latter treatments \emph{and the spatial
fluctuations of Kondo temperatures of the electronic Griffiths phase
we find in our approach}. We defer the discussion of this problem to a
future publication.

\begin{acknowledgments}
We thank M. J. Rozenberg for useful discussions. We also thank
C. H. Booth for providing us his XAFS data. This work was supported by
FAPESP through grants 99/00895-9 (MCOA), 98/12741-3 (EM), 01/00719-8
(EM), by CNPq through grant 301222/97-5 (EM), and by the NSF through
grants DMR-9974311 and NSF-0234215 (VD).
\end{acknowledgments}
\appendix

\section{Brief description of the numerical method}

\label{sec:numerics}

The set of stochastic equations defined
in~(\ref{eq:siteaction}-\ref{eq:bath2}) must be solved in two
steps. First, action (\ref{eq:siteaction}-\ref{eq:siteactionhyb}) is
solved with the bath function with $z-1$ nearest neighbors defined in
(\ref{eq:bath2}).  This determines self-consistently the distribution
of local conduction electron Green's functions with one nearest
neighbor removed $G_{cl}^{loc\left(j\right)}\left(i\omega
_{n}\right)$.  Next, the same action
(\ref{eq:siteaction}-\ref{eq:siteactionhyb}) is solved, this time with
the bath function in (\ref{eq:bath}), constructed from the previously
determined $G_{cl}^{loc\left(j\right)}\left(i\omega _{n}\right)$. This
step involves no self-consistency and yields the distribution of
$G_{cj}^{loc}\left(i\omega _{n}\right)$ as output. Since the latter
bath function is a sum over $z$ nearest neighbors its statistical
fluctuations are reduced compared to the former one. Thus,
$G_{cj}^{loc}\left(i\omega _{n}\right)$ is more narrowly distributed
than $G_{cl}^{loc\left(j\right)}\left(i\omega _{n}\right)$.  Yet, we
expect the qualitative behavior to be the same. We have therefore
focused on the first step of the procedure only.

For a given impurity solver, this disordered Bethe lattice problem was
solved for $z=3$, by sampling the distribution of
$G_{cj}^{loc\left(l\right)}\left(i\omega _{n}\right)$ from an ensemble
of $N$ sites, as proposed originally in Ref.~\onlinecite{abouetal}.
We have generally used $N=70-100$ since we have checked that the
results do not change by taking $N=200$. We have thus determined the
distribution of various local properties.

The equations were solved on a discrete mesh along the Matsubara
axis. The mesh is set by the Matsubara frequencies in the perturbative
treatment at finite temperatures and by an arbitrary finite discrete
mesh in the slave boson mean field theory at $T=0$ (up to 32000
points). The choice of the imaginary frequency axis is due to a
greater numerical stability. When the disorder is strong, the various
Green's functions show large fluctuations. However, these are much
more pronounced on the real frequency axis, where they give rise to
several peaks and gaps.

The slave boson treatment consists in finding the two mean field
parameters $q$ and $\epsilon _{f}$ by solving the set of two
non-linear Eqs.~(\ref{eq:sb1b}-\ref{eq:sb2b}).  For that, we used the
Powell hybrid method. The integrals were calculated with standard
adaptive quadrature routines. Since we used a finite frequency mesh,
it was important to extrapolate the value of $G_{f}^{qp}\left(i\omega
\right)$ with the asymptotic form $1/i\omega $ for values of $\omega $
greater than the largest mesh value.  In this fashion, a wide range of
Kondo temperatures is covered, going down to almost machine precision
in the clean metallic case. When disorder is present a given impurity
problem may not have a solution even at $T=0$, because the strong
spatial fluctuations may cause the local DOS to vanish at the Fermi
level. In this case, there are two possible regimes for the impurity,
which have been carefully analyzed.\cite{vladgabiSCKondo,SCKE1,SCKE2}
The analysis shows that there is a critical coupling constant $V_{c}$
such that the ground state is a singlet for $V>V_{c}$ (the so-called
{}``strong coupling Kondo effect''), whereas the local moment remains
unquenched if $V<V_{c}$. When a solution could not be found, this
corresponded to either a free spin ($V<V_{c}$) or a Kondo temperature
which is smaller than the smallest value we can reach with our
numerical code. In either case, we set $q=0$, effectively decoupling
the free moment from the rest of the lattice. Yet, we were still able
to span several decades of energy scales.

In the perturbative treatment, the solution of each impurity problem
is found by solving a set of two non-linear equations for $n$ and
$\tilde{\mu }$, which is defined by Eqs. (\ref{eq:nf}) and
(\ref{eq:neqn0}). As in the slave boson treatment we used the Powell
hybrid method. The calculation of the second order correction for the
self-energy involves Fourier transforms as, according to
Eq. (\ref{eq:2ordselfen}), it has a simpler form in imaginary time
rather than in frequency space. For this, we used the Fast Fourier
Transform algorithm.\cite{numrec}

\end{document}